\documentclass[longauth,traditabstract]{aa}  
\usepackage{graphicx}
\usepackage{natbib}
\bibpunct[; ]{(}{)}{,}{a}{}{;} 
\usepackage{lscape}
\usepackage{longtable}
\usepackage{rotating}
\usepackage{pdflscape}
\usepackage{tablefootnote}
\usepackage{txfonts}

\usepackage{float}
\usepackage{xcolor}

\newcommand{\teff}{$T_{\rm eff}$}
\newcommand{\logg}{$\log g$}
\newcommand{\vsini}{$v\sin i$}

\newcommand{\vinf}{$v_\infty$}

\usepackage[normalem]{ulem}

\begin{document} 

   \title{X-Shooting ULLYSES: massive stars at low metallicity}
\subtitle{I.\ Project Description\thanks{Based on observations collected at the European Southern Observatory under ESO programme 106.211Z.001.}}

\titlerunning{X-Shooting ULLYSES -- I. Project Description}
\authorrunning{Vink et al.}

  \author{Jorick S. Vink\inst{\ref{instAOP}}
   \and
      {A. Mehner\inst{\ref{eso}}} 
      \and 
      {P. A. Crowther\inst{\ref{Shef}}}
    \and
    {A. Fullerton\inst{\ref{STScI}}}
    \and
        {M. Garcia\inst{\ref{cabt}}}
      \and
      {F. Martins\inst{\ref{lupm}}} 
      \and
      {N. Morrell\inst{\ref{lco}}}
       \and
     {L.M. Oskinova\inst{\ref{UP}}}
     \and
      {N. St-Louis\inst{\ref{udem}}}
      \and
      {A. ud-Doula\inst{\ref{PSU}}}
     \and
      {A.A.C.\ Sander\inst{\ref{ari}}}
   \and
      {H. Sana\inst{\ref{KUL}}}
      \and
      {J.-C. Bouret\inst{\ref{LAM}}}
       \and
     {B.~Kub\'atov\'a\inst{\ref{asu}}}   
    \and
      {P. Marchant\inst{\ref{KUL}}}
      \and
      {L.P. Martins\inst{\ref{unicid}}}  
    \and
    {A. Wofford\inst{\ref{unam}}}
    \and
      {J. Th. van Loon\inst{\ref{Keele}}}
      \and
      {O. Grace Telford\inst{\ref{rut}}}
      \and             
      {Y. G\"otberg\inst{\ref{carnegie}}}
       \and  
        {D.M. Bowman\inst{\ref{KUL}}}
    \and
      {C. Erba\inst{\ref{ETSU}}}
      \and
      {V.M.~Kalari\inst{\ref{noir}}}
    \and    
      {M. Abdul-Masih\inst{\ref{eso}}}
      \and
      {T. Alkousa\inst{\ref{Shef}}}
   \and
      {F. Backs\inst{\ref{API}}}
   \and
      {C.L. Barbosa\inst{\ref{fei}}}
   \and
   {S.R. Berlanas\inst{\ref{UA},\ref{Keele}}}
   \and
   {M. Bernini-Peron \inst{\ref{ari}}}
   \and
   {J.M. Bestenlehner\inst{\ref{Shef}}}
   \and 
      {R. Blomme\inst{\ref{rob}}}
   \and
      {J. Bodensteiner\inst{\ref{eso_garching}}}
    \and
      {S.A. Brands\inst{\ref{API}}}
      \and
      {C.J. Evans\inst{\ref{STScI}}}
    \and
      {A. David-Uraz \inst{\ref{Howard}, \ref{NASA}}}
    \and
    { F.A. Driessen \inst{\ref{KUL}, \ref{Leiden}}}
    \and
    {K. Dsilva \inst{\ref{KUL}}}
    \and
              {S. Geen\inst{\ref{API}, \ref{Leiden}}}
              \and
      {V.M.A.G\'omez-Gonz\'alez \inst{\ref{UP}}}      
    \and
    {L. Grassitelli \inst{\ref{aifa}}} 
    \and
       {W.-R. Hamann \inst{\ref{UP}}}  
       \and
       {C. Hawcroft \inst{\ref{KUL}}}
    \and
       {A. Herrero \inst{\ref{IAC}}}
    \and
        {E.R. Higgins \inst{\ref{instAOP}}}
    \and    
       {D. John Hillier \inst{\ref{PITT}}}
    \and
       {R. Ignace \inst{\ref{ETSU}}}
    \and
       {A.G. Istrate \inst{\ref{IMAP}}}
    \and
      {L.~Kaper\inst{\ref{API}}}
    \and
    {N.D. Kee\inst{\ref{nso}}}
    \and
    {C. Kehrig\inst{\ref{iaa}}}
    \and
    {Z. Keszthelyi\inst{\ref{API}, \ref{NAOJ}}}
    \and
    {J. Klencki\inst{\ref{eso_garching}, \ref{MPA}}}
    \and
    {A. de Koter\inst{\ref{API}, \ref{KUL}}}
    \and
     {R.~Kuiper\inst{\ref{ude}}}
     \and
      {E. Laplace\inst{\ref{hits}}}
      \and
      {C.J.K. Larkin\inst{\ref{ari}}}
    \and
      {R. R. Lefever\inst{\ref{ari}}}
 \and 
      {C. Leitherer\inst{\ref{STScI}}}
      \and
      {D.J. Lennon\inst{\ref{IAC},\ref{lag}}}
    \and 
      {L. Mahy\inst{\ref{rob}}}
    \and
      {J. Ma\'{\i}z Apell\'aniz\inst{\ref{CAB}}}
    \and
      {G. Maravelias\inst{\ref{noa}, \ref{forth}}}
        \and
      {W. Marcolino\inst{\ref{ufrj}}} 
    \and
      {A. F. McLeod\inst{\ref{durham1}, \ref{durham2}}}
      \and
      {S.E. de Mink\inst{\ref{MPA}, \ref{API}}}
    \and
    {F. Najarro\inst{\ref{cabt}}}
    \and
    {M. S. Oey\inst{\ref{umich}}}
    \and
    {T.N. Parsons\inst{\ref{ucl}}}
    \and
      {D. Pauli\inst{\ref{UP}}}  
     \and
     {M.G. Pedersen\inst{\ref{kitp}}}
    \and
     {R.K. Prinja\inst{\ref{ucl}}}
     \and
      {V. Ramachandran\inst{\ref{ari}}}
     \and
      {M.C.~Ram\'irez-Tannus\inst{\ref{mpia}}}
    \and
     {G.N. Sabhahit \inst{\ref{instAOP}}}
     \and
      {A. Schootemeijer\inst{\ref{aifa}}}
\and
       {S. Reyero Serantes\inst{\ref{UP}}}  
    \and
      {T.\ Shenar\inst{\ref{API}}} 
    \and
    {G.S. Stringfellow\inst{\ref{casa}}}
    \and
      {N. Sudnik\inst{\ref{udem}}}  
    \and
      {F. Tramper\inst{\ref{KUL}}}
    \and
      {L. Wang\inst{\ref{ynao}}}
    }
   \institute{Armagh Observatory and Planetarium, College Hill, BT61 9DG Armagh, Northern Ireland \label{instAOP}\\ \email{jorick.vink@armagh.ac.uk}
   \and
   {Institute of Astronomy, KU Leuven, Celestijnenlaan 200D, B-3001 Leuven, Belgium \label{KUL}\\
   \email{dominic.bowman@kuleuven.be}}
   \and 
   {Dept of Physics \& Astronomy, University of Sheffield, Hounsfield Road, Sheffield, S3 7RH, UK \label{Shef}}\\ \email{paul.crowther@sheffield.ac.uk}
   \and
  {Penn State Scranton, 120 Ridge View Drive, Dunmore, PA 18512, USA \\
  \label{PSU}
\email {asif@psu.edu}}
    \and
    Centre for Extragalactic Astronomy, Department of Physics, Durham University, South Road,  Durham DH1 3LE, UK \\ \label{durham1}
    \email{anna.mcleod@durham.ac.uk}
\and
Institute for Computational Cosmology, Department of Physics, University of Durham, South Road, Durham DH1 3LE, UK  \label{durham2}
\and
  {Department of Physics and Astronomy, East Tennessee State University, Johnson City, TN 37614, USA \\
  \label{ETSU}
\email {christi.erba@gmail.com}}
\and
{Department of Physics and Astronomy \& Pittsburgh Particle Physics, Astrophysics and Cosmology Center (PITT PACC), University of Pittsburgh, 3941 O'Hara Street,  Pittsburgh, PA 15260, USA \\
\label{PITT} \email{hillier@pitt.edu.}}
   \and
 {Astronomick\' y \' ustav, Akademie v\v ed \v Cesk\' e republiky, CZ-251 65 Ond\v rejov, Czech Republic \\
\label{asu} \email{brankica.kubatova@asu.cas.cz}}  
   \and
   {Instituto de Astrofisica de Canarias, 38200, La Laguna, Tenerife, Spain \\
   \label{IAC}}
   \and
   {Dpto. Astrofisica, Universidad de La Laguna, 38 205 La Laguna, Tenerife, Spain\\
   \label{lag}}
   \and
   {Argelander Institute f{\"u}r Astronomie der Universit{\"a}t Bonn, Auf dem H{\"u}gel 71, 53121 Bonn, Germany \\ \label{aifa}}
\email {luca@astro.uni-bonn.de}
   \and 
  {LUPM, Universit\'e de Montpellier, CNRS, Place Eug\`ene Bataillon, F-34095 Montpellier, France \\
  \label{lupm}
\email {fabrice.martins@umontpellier.fr}}
   \and 
  {ESO - European Organisation for Astronomical Research in the Southern Hemisphere, Alonso de Cordova 3107, Vitacura, Santiago de Chile, Chile \\
  \label{eso}
\email {amehner@eso.org, mabdulm@eso.org}}  
   \and 
  {ESO - European Organisation for Astronomical Research in the Southern Hemisphere, Karl-Schwarzschild-Str. 2, 85748 Garching b. München, Germany \\
  \label{eso_garching}
  \email{julia.bodensteiner@eso.org}}
     \and
  {Las Campanas Observatory, Carnegie Observatories, Casilla 601, La Serena, Chile \\
  \label{lco} \email{nmorrell@carnegiescience.edu}}
  \and
  {Anton Pannekoek Institute for Astronomy, Universiteit van Amsterdam, Science Park 904, 1098 XH Amsterdam, The Netherlands \\
  \label{API} \email{s.t.geen@uva.nl}}
  \and
  {Leiden Observatory, Leiden University, NL-2300 RA Leiden, the Netherlands \\ \label{Leiden}}
  \and
  {Center for Computational Astrophysics, Division of Science, National Astronomical Observatory of Japan, 2-21-1, Osawa, Mitaka, Tokyo 181-8588, Japan  \\
  \label{NAOJ} \email{zsolt.keszthelyi@nao.ac.jp}}
   \and
  {Zentrum für Astronomie der Universit\"{a}t Heidelberg,  Rechen-Institut, M\"{o}nchhofstr. 12-14, 69120 Heidelberg}\\
   \label{ari} \email{andreas.sander@uni-heidelberg.de}
   \and
   {NAT - Universidade Cidade de Sao Paulo, Rua Galvao Bueno, 868, S\~ao Paulo, Brazil\\
   \label{unicid} \email{lucimara.martins@cruzeirodosul.edu.br}}
  \and
  {Department of Physics and Astronomy, University College London,Gower Street,London WC1E 6BT,UK \\ 
  \label{ucl}}
  \and
  {Fakult\"{a}t für Physik, Universit\"{a}t Duisburg-Essen, Lotharstraße 1, 47057 Duisburg, Germany \\
  \label{ude} \email{rolf.kuiper@uni-due.de}}
    \and
  {Gemini Observatory/NSF’s NOIRLab, Casilla 603, La Serena, Chile \\
  \label{noir} \email{venu.kalari@noirlab.edu}}
  \and
  {University of Michigan, Department of Astronomy, 323 West Hall, Ann Arbor, MI   48109, USA\\
  \label{umich}}
  \and
  {Institute for Physics and Astronomy, University Potsdam, D-14476 Potsdam, Germany\\
  \label{UP}  
  \email{lida@astro.physik.uni-potsdam.de}}
  \and
  {Kavli Institute for Theoretical Physics, Kohn Hall, University of California, 
Santa Barbara, CA 93106, USA\\
  \label{kitp}  
  \email{mgpedersen@kitp.ucsb.edu}}
  \and
  {Centro de Astrobiolog\'{\i}a (CAB), CSIC-INTA. Campus ESAC. C. bajo del castillo s/n. E-28\,692 Madrid, Spain\\
  \label{CAB}  
  \email{jmaiz@cab.inta-csic.es}}
  \and 
  {Heidelberger Institut f\"{u}r Theoretische Studien, Schloss-Wolfsbrunnenweg 35, 69118 Heidelberg, Germany\\
  \label{hits}\email{eva.laplace@h-its.org}}
  \and
  {Center for Astrophysics and Space Astronomy, University of Colorado Boulder, Boulder, CO 80309-0389,USA \\ \label{casa}}
 \and 
 {Aix Marseille Univ, CNRS, CNES, LAM, Marseille, France\label{LAM}\email{Jean-Claude.Bouret@lam.fr}}
 \and
 {D\'epartement de physique, Universit\'e de MOntr\'eal, Campus MIL, 1375 Th\'er`ese-Lavoie-Roux, Montr\'eal (QC), H2V 0B3, Canada \label{udem}\email{nicole.st-louis@umontreal.ca} }
 \and
 {Royal Observatory of Belgium, Avenue circulaire/Ringlaan 3, B-1180 Brussels, Belgium \\ \label{rob}\email{laurent.mahy@oma.be}}
  \and
 {Rutgers University, Department of Physics and Astronomy, 136 Frelinghuysen Road, Piscataway, NJ 08854, USA \\ \label{rut}\email{grace.telford@rutgers.edu} }
 \and
 {Lennard-Jones Laboratories, Keele University, ST5 5BG, UK\\
 \label{Keele}\email{j.t.van.loon@keele.ac.uk}}
  \and
 {The Observatories of the Carnegie Institution for Science, 813 Santa Barbara Street, CA-91101 Pasadena, USA\\
 \label{carnegie}\email{ygoetberg@carnegiescience.edu}}
 \and
 {Centro de Astrobiolog\'{\i}a (CAB), CSIC-INTA. Ctra. Torrej\'on a Ajalvir km 4., 28850, Torrej\'on de Ardoz, Madrid, Spain\\
 \label{cabt}
 \email{mgg@cab.inta-csic.es,najarro@cab.inta-csic.es}}
 \and
 {Departamento de Física Aplicada, Universidad de Alicante, E-03\,690, San Vicente del Raspeig, Alicante, Spain\\
 \label{UA}
 \email{sara.rb@ua.es}}
 \and
 {Department of Physics and Astronomy, Howard University, Washington, DC 20059, USA\\
 \label{Howard}
 \email{alexandre.daviduraz@howard.edu}}
 \and
 {Observatório do Valongo, Universidade Federal do Rio de Janeiro, Ladeira Pedro Antônio 43, Rio de Janeiro, CEP 20080-090, Brazil\\
 \label{ufrj}
 \email{wagner@ov.ufrj.br}}
  \and
 {Instituto de Astrof\'{\i}sica de Andaluc\'{\i}a - CSIC, Glorieta de la Astronom\'{\i}a s/n, 18008, Granada, Spain\\
 \label{iaa}}
 \and
 {Center for Research and Exploration in Space Science and Technology, and X-ray Astrophysics Laboratory, NASA/GSFC, Greenbelt, MD 20771, USA\\
 \label{NASA}}
 \and
 {Nicolaus Copernicus Astronomical Centre of the Polish Academy of Sciences, Bartycka 18, 00-716 Warszawa, Poland\\
 \label{CAMK}
 \email{snata.astro@gmail.com}}
 \and
 {Space Telescope Science Institute, 3700 San Martin Dr, Baltimore, MD 21218, USA\\
 \label{STScI}
 \email{leitherer@stsci.edu}}
 \and
 {Max Planck Institut f\"ur Astronomie,  K\"onigstuhl 17, 69117 Heidelberg, Germany\\
 \label{mpia}}
 \and
 {Max Planck Institut f\"ur Astrophysik, Karl-Schwarzschild-Strasse 1, 85741 Garching, Germany\\
 \label{MPA}}
\and
{Centro Universit\'ario FEI, Dept. de F\'{\i}sica.
Av. Humberto Alencar de Castelo Branco, 3972
S\~ao Bernardo do Campo - SP, CEP 09850-901, Brazil\\
\label{fei}
\email{cbarbosa@fei.edu.br}}
\and
{IAASARS, National Observatory of Athens, GR-15236, Penteli, Greece \\ \label{noa}
   \email{maravelias@noa.gr}}
\and
{Department of Astrophysics/IMAPP, Radboud University Nijmegen, P.O. Box 9010, 6500 GL    Nijmegen, The Netherlands \\ \label{IMAP}
}
\and
{Instituto de Astronom\'ia, Universidad Nacional Aut\'onoma de M\'exico, Unidad Acad\'emica en Ensenada, Km 103 Carr. Tijuana$-$Ensenada, Ensenada, B.C., C.P. 22860, M\'exico \\ \label{unam}}
\and
{National Solar Observatory, 22 Ohi`a Ku St, Makawao, HI 96768, USA\\ \label{nso}}
\and
{Institute of Astrophysics, FORTH, GR-71110, Heraklion, Greece \\ \label{forth}
   \email{gmaravel@ia.forth.gr}}
 \and
 {Yunnan Observatories, Chinese Academy of Sciences, Kunming 650216, Yunnan, China \\ \label{ynao}
 \email{wangluqian@ynao.ac.cn}}
}

 \date{Received Mar 2.71828182845, 2023; accepted May 3.14159265358979, 2023}

 
  \abstract
  {Observations of individual massive stars, super-luminous supernovae, gamma-ray bursts, and gravitational-wave events involving spectacular black-hole mergers, indicate that the low-metallicity Universe is fundamentally different from our own Galaxy. Many transient phenomena will remain enigmatic until we achieve a firm understanding of the physics and evolution of massive stars at low metallicity ($Z$). 
   The {\it Hubble Space Telescope\/} has devoted 500 orbits to observe $\sim$250 massive stars at low $Z$ in the ultraviolet (UV) with the {\it COS\/} and {\it STIS\/} spectrographs under the ULLYSES program. 
   The complementary ``X-Shooting ULLYSES'' (XShootU) project provides enhanced legacy value with high-quality optical and near-infrared spectra obtained with the wide-wavelength coverage X-shooter spectrograph at ESO's Very Large Telescope. 
   We present an overview of the XShootU project, showing that combining ULLYSES UV and XShootU optical spectra is critical for the uniform determination of stellar parameters such as effective temperature, surface gravity, luminosity, and abundances, as well as wind properties such as mass-loss rates in function of $Z$. As uncertainties in stellar and wind parameters percolate into many adjacent areas of Astrophysics, the data and modelling of the XShootU project is expected to be a game-changer for our physical understanding of massive stars at low $Z$.
   To be able to confidently interpret {\it James Webb Space Telescope\/} (JWST) spectra of the first stellar generations, the individual spectra of low $Z$ stars need to be understood, which is exactly where XShootU can deliver.
   }


   \maketitle
%

\section{Introduction}

We find ourselves amidst a scientific revolution: gravitational wave (GW) observatories will soon be detecting black hole (BH) mergers as frequently as once per day.
To interpret these events, we need to comprehend massive stars in low metallicity ($Z$) environments \citep{2020ApJ...892L...3A}. 
This is also crucial for other fields of Astrophysics, including feedback processes (e.g., \citealt{2013A&A...558A.134D}), star formation, interstellar medium (ISM) physics, supernovae (SNe), and cosmology. 
To enable progress in these research areas, we need to uniformly sample the relevant parameter space for massive OB stars, including spectral type, luminosity class, and metallicity ($Z$). 

\begin{figure*}
    \centering
    \includegraphics[width=\textwidth]{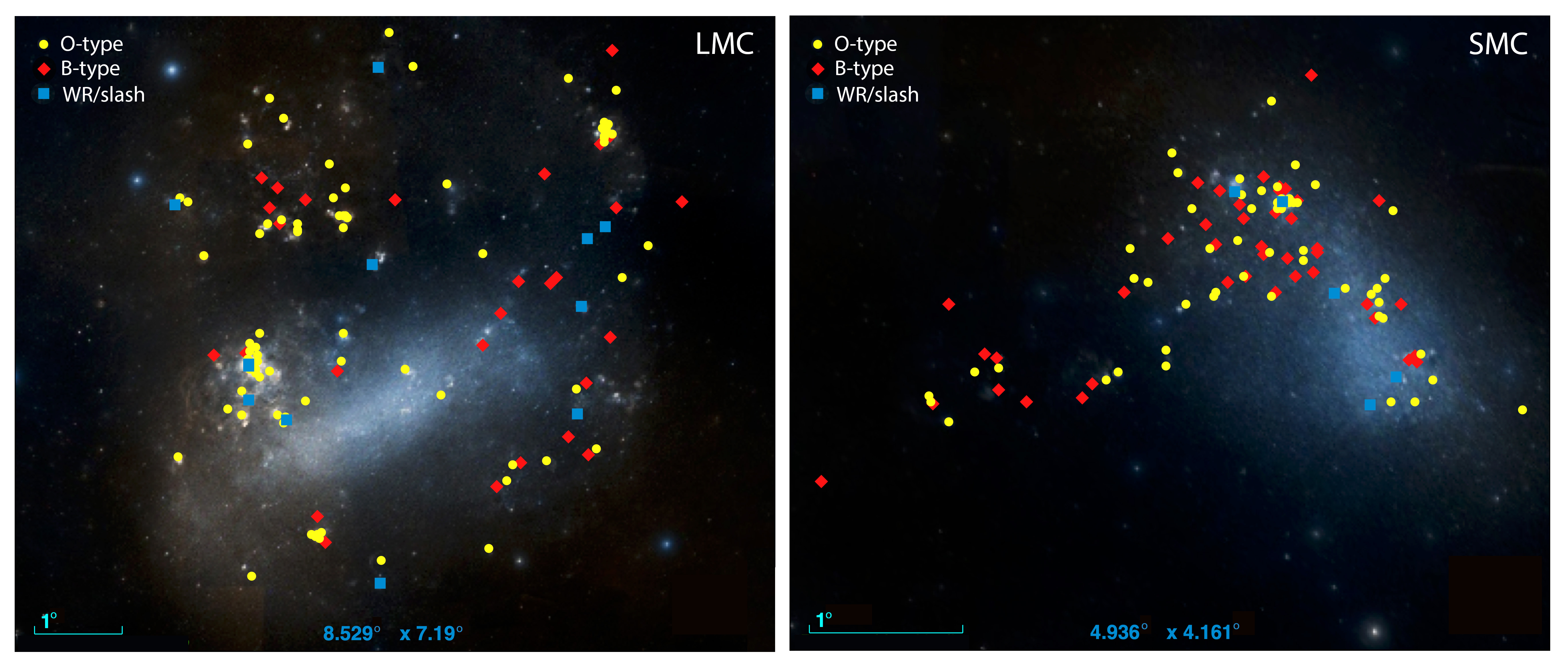}
    \caption{Positions of the ULLYSES/XShootU sources in the LMC (left) and SMC (right). Yellow dots are O-type stars, red diamonds are B-type stars, and blue squares are Wolf-Rayet (WR) and WR-like Of/WR "slash" stars. We note that the two images have different spatial scales. This figure was made with the Aladin Sky Atlas \citep{2000A&AS..143...33B}; the background consists of DSS2 colour images. \label{fig:skymap}}
\end{figure*}

\begin{figure}
    \centering
    \includegraphics[width=0.49\textwidth]{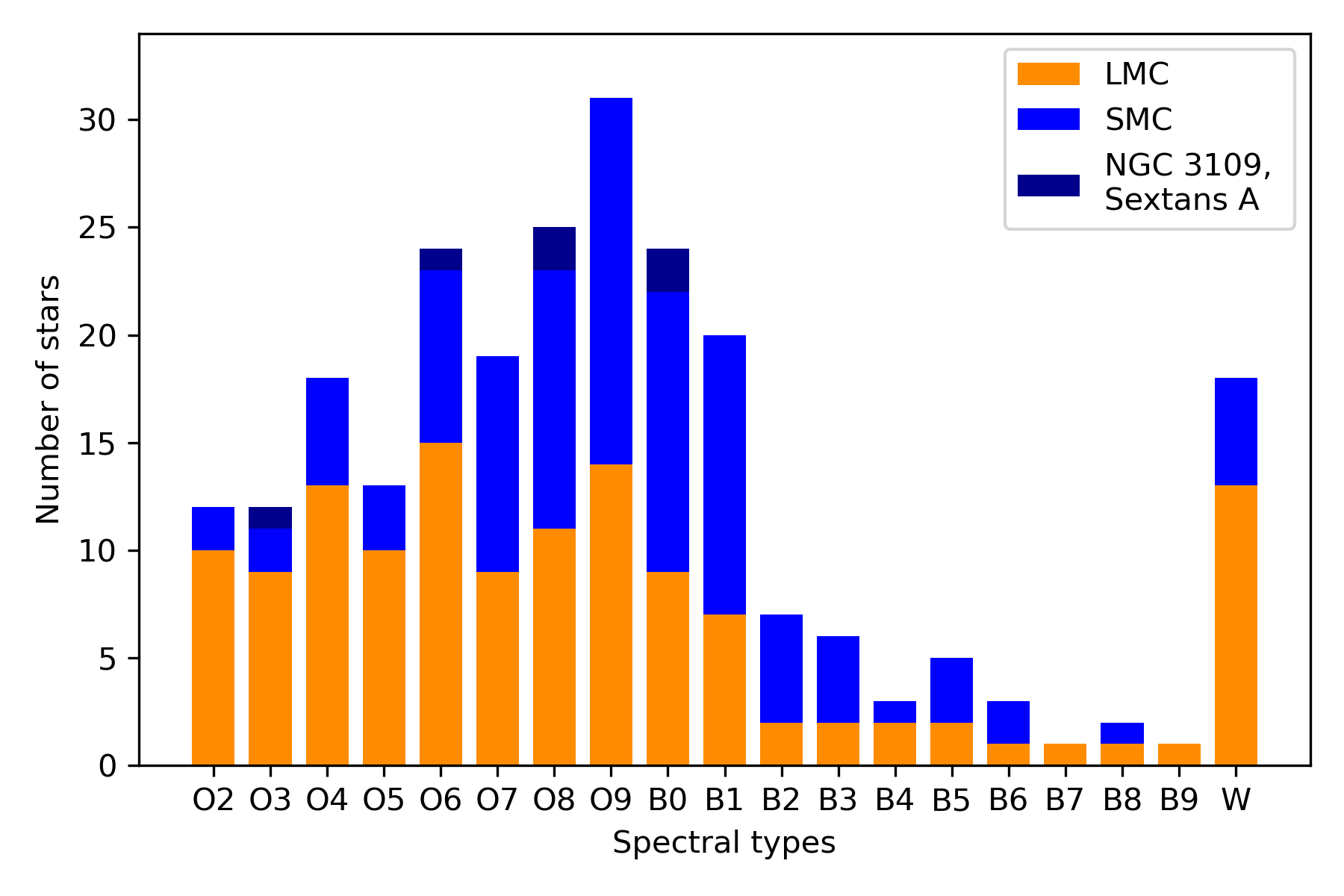}
    \caption{Distribution of the spectral types in ULLYSES. For the 10 known binaries in the sample only the primary component is accounted for. The category labeled `W' includes Wolf-Rayet (WR) and WR-like "slash" stars.}
        \label{f:sptype}
\end{figure}

The {\it Hubble Space Telescope} ({\it HST\/}) has dedicated 1000 Orbits to the Director's Discretionary Time project ``Ultraviolet Legacy Library of Young Stars as Essential Standards'' (ULLYSES; \citealt{2020RNAAS...4..205R}),\footnote{\url{https://ullyses.stsci.edu}}, making this the largest HST program ever conducted.
ULLYSES compiles an ultraviolet (UV) spectroscopic Legacy Atlas of about 250 OB stars in low $Z$ regions.\footnote{The ULLYSES program is also compiling high-quality far-UV, near-UV, and optical spectra of young, low-mass T Tauri stars in our Galaxy.} 
Due to their proximity, the Large and Small Magellanic Clouds (LMC, SMC) are the best low-$Z$ laboratories for massive star studies, with respectively 50\% and 20\% $Z_{\odot}$. They are ideal to study spatially resolved populations of low-$Z$ massive stars to make a leap towards understanding the Early Universe. As a pilot study, several stars at sub-SMC metallicities (in Sextans~A and NGC~3109; $\sim$10\% $Z_{\odot}$) are also included. 
The aim is to uniformly cover all spectral sub-types O2$ - $B9 and to observe all luminosity classes with spectral types O2$ - $B1.5, for both LMC and SMC metallicity, leading to $\sim 250$ stars (Figures ~\ref{fig:skymap} and \ref{f:sptype}). 

\begin{figure*}
    \centering
    \includegraphics[angle=0,width=0.49\textwidth]{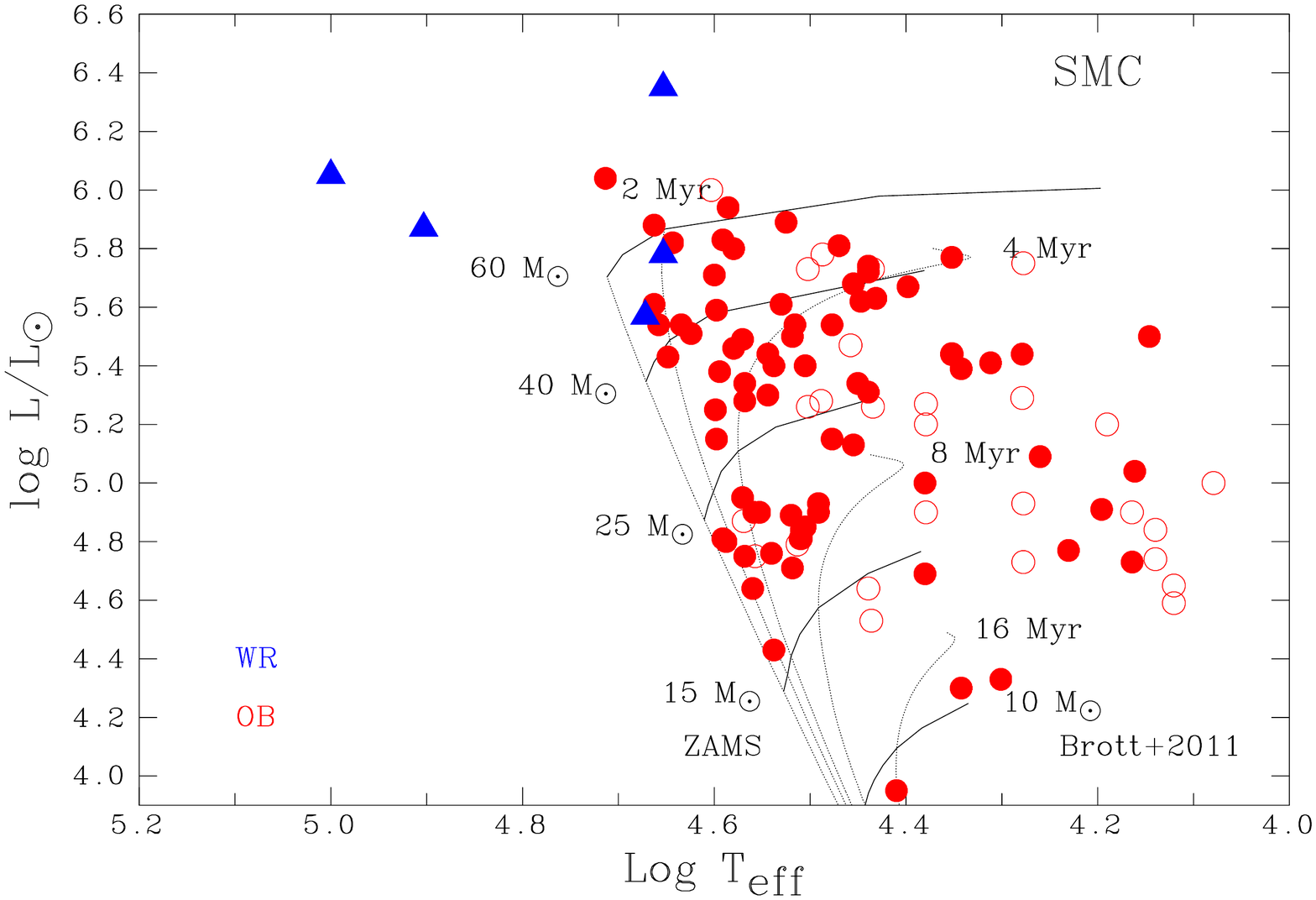}
    \includegraphics[angle=0,width=0.49\textwidth]{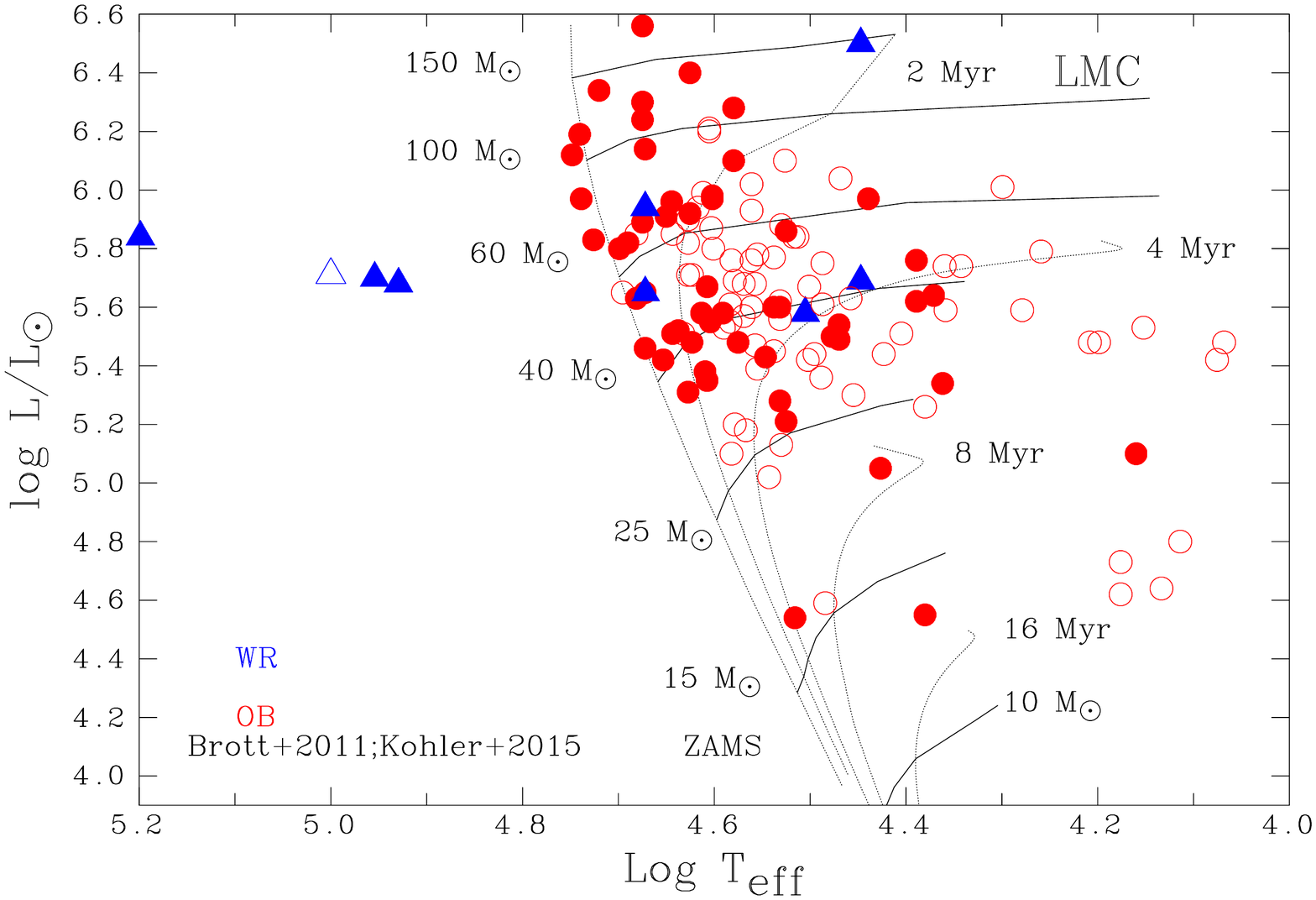}
    \caption{HR diagrams of the ULLYSES SMC and LMC targets. Stellar parameters are based on contemporary literature from Table\,\ref{table:params} (filled symbols) or spectral type calibrations (open symbols). For known binary/multiple systems, only primaries are indicated. Calibrations used are \citet{2013A&A...558A.134D} for O-type stars in both galaxies, \citet{2019A&A...626A..50D,2004A&A...417..217T,2005A&A...434..677T} for SMC B-type stars and \citet{2018A&A...615A.101D,2017A&A...603A..91G,2015A&A...575A..70M,2017AJ....154..102U} for LMC B-type stars. Evolutionary tracks (solid lines) and isochrones (dotted lines) for non-rotating massive stars at 0.5 $Z_{\odot}$ and 0.2~$Z_{\odot}$ are from \citet{2011A&A...530A.115B}, supplemented by tracks for very massive stars in the LMC from \citet{2015A&A...573A..71K}. }
    \label{SMC-LMC-HRD}
\end{figure*}

Although massive stars emit the bulk of their light at UV wavelengths, the optical region remains the cornerstone of spectroscopic analysis studies. 
The UV regime is powerful to determine the iron (Fe) abundance and to obtain information on wind parameters, such as the terminal velocity ($v_{\infty}$). The optical regime is crucial to determine the basic stellar parameters, such as effective temperature ($T_{\rm eff}$), surface gravity (\logg), and abundances \citep{Hillier20,Simon20,Brands:2022} 
Key information on wind clumping and mass-loss rates ($\dot{M}$) only become reliable when optical and near-infrared (NIR) observations are added. Knowledge of the NIR regime is also critical for observations with instrumentation at the {\it James Webb Space Telescope\/} ({\it JWST\/}) and the Extremely Large Telescopes, which will predominately shift our focus to longer wavelengths.  
Despite the great potential of ULLYSES to transform our knowledge of massive stars, this Legacy dataset is not complete without observations in the optical and NIR regimes. Thus, the XShootU\footnote{https://massivestars.org/xshootu/} project was conceived to obtain complementary high quality spectra of the ULYSSES targets with X-shooter at ESO's Very Large Telescope \citep{2011A&A...536A.105V}. 

The optical Large VLT-Flames Survey of massive stars (PI: S.J.\ Smartt) and its successor, the VLT-Flames Tarantula survey VFTS (PI: C.J.\ Evans), tackled many science questions including the $Z$-dependence of stellar wind mass-loss rates \citep{2007A&A...473..603M} and the rotation velocities of massive stars \citep{2013A&A...560A..29R}. The surface nitrogen (N) abundance of most massive stars in the MCs seemed to be consistent with theoretical predictions, but a significant fraction of stars ($20\%-40$\% depending on sample and metallicity) showed chemical enrichment that is either too strong or too weak (e.g., \citealt{2008ApJ...676L..29H,2010A&A...517A..38P,2014A&A...565A..39M,2017A&A...600A..82G}). 

The absolute mass-loss rates of massive OB and Wolf-Rayet (WR) stars are still uncertain (e.g., \citealt{2019A&A...632A.126S,2020MNRAS.491.4406S,2019A&A...625A.104R,Marcolino22,Rickard22}). According to evolutionary models, the bulk of the mass loss could occur during the B-supergiant phase rather than during the preceding O-star phase \citep{2014A&A...564A..30G}.
The $Z$ dependence of mass-loss behaviour in this cooler regime is highly complex, involving various mass-loss discontinuities as a function of temperature (bi-stability jumps; \citealt{2016MNRAS.458.1999P}), and is critical in predicting BH masses as a function of $Z$ \citep{2010ApJ...714.1217B} as well as GW mergers \citep{2016A&A...596A..58K}.

The combined UV and optical XShootU project was motivated to address these science questions as well as a large variety of additional questions concerning massive stars at low $Z$. 
The project will derive accurate stellar and wind parameters, such as effective temperatures, luminosities, gravities, abundances, and mass-loss rates. 
This will establish whether mass-loss rates are decreasing with lower $Z$,
as predicted \citep{2001A&A...369..574V,Kudr02} and empirically supported for relatively small LMC and SMC VLT-Flames survey samples \citep{2007A&A...473..603M,2019A&A...625A.104R}. 
Moreover, feedback parameters involving wind momenta, wind kinetic energy, and ionising fluxes are key ingredients for building the next generation of spectral population synthesis models that may be applied to extra-galactic surveys, such as CLUES \citep{Sirressi22}, CLASSY \citep{Berg22}, the {\it HST\/} spectroscopic survey of star-forming galaxies in the Local Universe, and future projects. 
Bright early-type stars are also excellent probes of ISM conditions \citep{2013A&A...550A.108V}.
We expect many spin-off projects using XShootU and ULLYSES data, including the derivation of the extinction law for which X-shooter's wide spectral range is particularly useful.

In this work, we present the XShootU project. The science requirements are described, as well as  
initial results on data reduction and data analysis processes. 
We show how the legacy spectroscopic data-set of ULLYSES and XShootU can increase our knowledge of massive stars at low $Z$. The organisation of the XShootU collaboration is described in the Appendix.
Already published data (Table\,\ref{table:params}, plotted in Fig.\,\ref{SMC-LMC-HRD}) may naively suggest that properties of LMC and SMC stars are known, but these pre-ULLYSES results have been derived for relatively small and heterogeneous data-sets, and gaps are evident. To make matters worse, spectral analyses to derive the stellar properties have also been heterogeneous, as different authors have not only used different tools, distances, and baseline abundances, but also different wavelength ranges. 

To make progress, not only the spectroscopic data sets need to be uniform -- as provided by ULLYSES and XShootU -- but so does the spectral analysis approach. To give an example, to determine the $\dot{M}-Z$ relationship not only require accurate mass-loss rate determinations, but also reliable stellar parameters, such as luminosities, to compare $\dot{M}$ from one star with one set of stellar properties in one particular galaxy to the mass-loss rate from another star in another galaxy. A uniform data and analysis approach is at the heart of the XShootU project.

\section{XShootU science requirements}
\label{s:req}

\begin{figure}
    \centering
    \includegraphics[width=0.95\columnwidth]{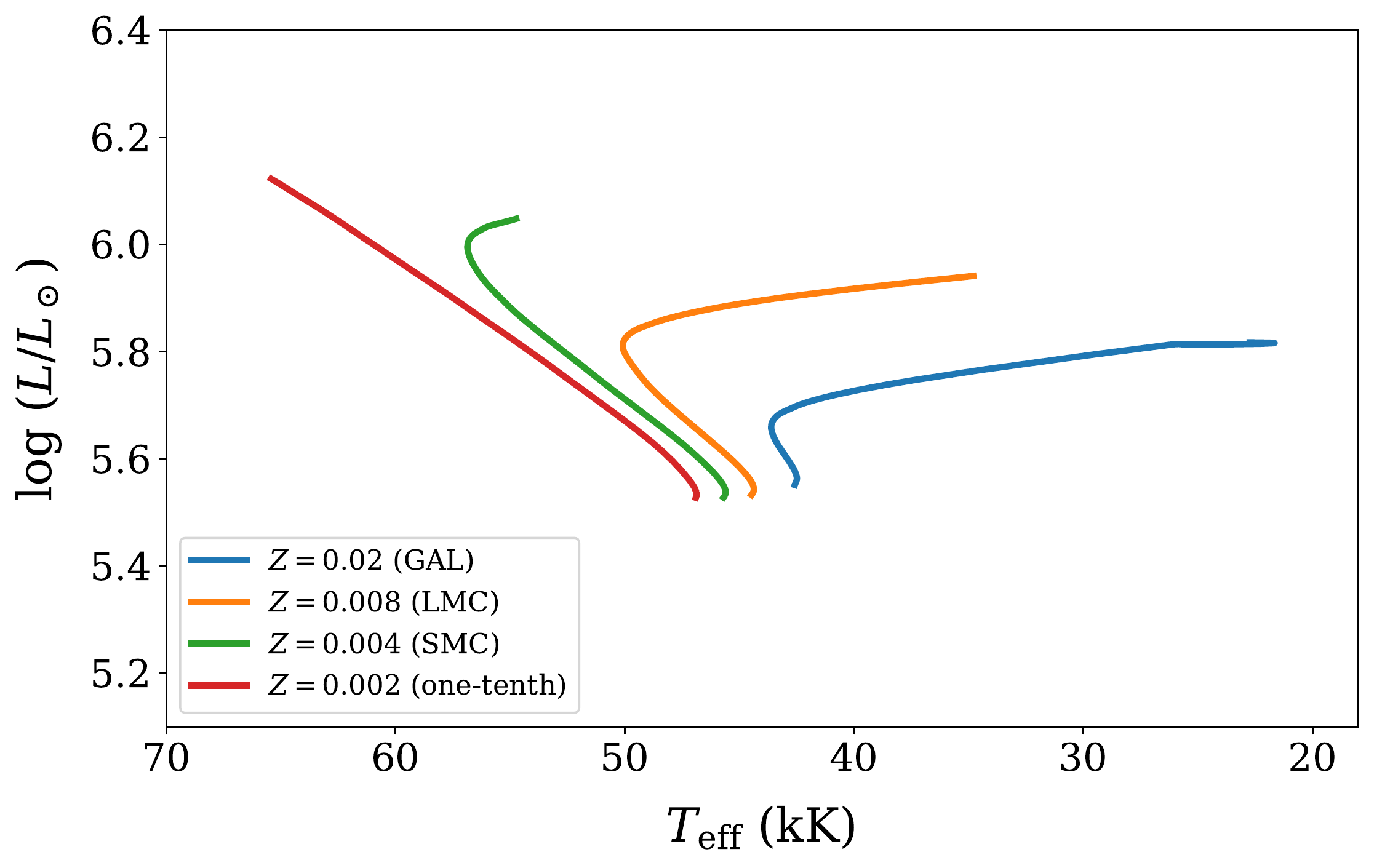}
     \caption{Main-sequence MESA stellar evolution models of a rapidly rotating ($v\sin{i} = 550$\,km/s) 50\,$M_{\odot}$ star for a range of metallicities. The Galactic and LMC model show traditional redwards evolution, while the SMC and even lower $Z$ (1/10th solar) models start to show bluewards chemical  
     homogeneous evolution. 
     The models employ Vink et al. (2000, 2001) mass-loss rates and assume a moderate amount of core overshooting with a value of $\alpha_{\rm ov}$ of $0.335$, similar to Brott et al. (2011) models.}
        \label{f:mesa}
\end{figure}

In order to build better population synthesis models of massive stars in low-$Z$ environments, such as those at high redshift studied with JWST, we require (i) more complete spectral libraries, as well as (ii) more reliable stellar evolution models for low-$Z$ stars. The former involves the construction of more accurate model atmospheres, but the latter implies a better handle on the behaviour of wind mass loss over a multi-dimensional parameter space, including $Z$. The key line driver of the inner winds that sets $\dot{M}$ of massive OB stars is iron (Fe), while intermediate mass elements such as CNO dominate the outer winds, setting the terminal velocity
\citep{Vink99,Puls2000}. 
While high redshift galaxies may have different [$\alpha$/Fe] ratios compared to local low-$Z$ galaxies, non-solar [$\alpha$/Fe] ratios should have very little impact on the expected mass-loss rate, as long as one correctly interprets low-$Z$ as having low Fe 
contents (see for instance Table 5 in Vink et al. 2001 for conversions between O and Fe). 

One potential concern is whether the local low $Z$ LMC and SMC at 0.5\,$Z_{\odot}$ and 0.2\,$Z_{\odot}$ are sufficiently metal-poor to gain insight into low-$Z$ stellar evolution in high-redshift galaxies.
In order to make the case that the SMC indeed has a sufficiently low Fe-contents to provide key insights into the Early Universe, we showcase a number of MESA stellar evolution models (see WG\,5 in Appendix) in Fig.\,\ref{f:mesa}. The plot indicates that a rapidly rotating massive star at LMC metallicity still shows classical redwards evolution, just like in the Milky Way, but that already the 1/5th solar SMC metallicity is sufficiently low to undergo bluewards chemically homogeneous evolution (CHE), similar to even lower $Z$ stars at 1/10th solar. 
The Galactic model loses as much as a third of its initial 50\,$M_{\odot}$ mass already on the main-sequence, while the SMC and lower-$Z$ models lose of order 10\% or less. 
Moreover, while the Galactic and LMC models completely spin down during the main sequence (the Galactic model drops below the minimum observable $v\sin{i}$ value of 100\,km/s after 3\,Myrs, while the LMC star can delay this to 4.5\,Myrs), the SMC and lower $Z$ models hardly spin down at all. 
In fact, the SMC and lower $Z$ model evolve towards critical rotation, rather than away from it.
In other words, the SMC is an ideal test-bed for gaining an understanding of the physical difference between the high-$Z$ and the low-$Z$ Universe.  

The results displayed in Fig.\,\ref{f:mesa} may naively give the impression that stellar evolution is already well understood, but this is not the case, and the stellar evolution and population synthesis models are only as good as the input physics. In this parameter space those are predominately given by the assumed amounts of interior mixing and wind mass loss. It is commonly assumed that the only parameters setting the mass-loss rate are the stellar luminosity and the metallicity, but in the oft-used mass-loss recipe of \cite{Vink:2000,2001A&A...369..574V} parameters such as stellar mass, and effective temperature -- including the B supergiant regime below the bistability jump --  also play a crucial role. Therefore, in order to make progress on the accuracy of stellar evolution models at low $Z$ we firstly require large samples of wind parameters offered by the ULLYSES sample. 
Secondly, in order to test these in different parts of the HR diagram, the underlying stellar parameters also need to be robust. Thirdly, in order to test the role of rotational mixing for a range of metallicities we require stellar abundances. 

Starting with the third requirement, massive stars undergo H-buring via the CNO cycle, and in the first instance the core nitrogen (N) abundance is expected to increase by an order of magnitude at the expense of carbon (C) \citep{2011A&A...530A.115B,ekstrom+12}. 
Mixing can bring enhanced N to the surface, which is especially relevant for testing the physics of rotational mixing in stellar evolution models. Factors 2-10 in N enhancement and C depletion are realistically measurable from UV and optical spectroscopy as discussed in Sect.\,\ref{s:multi}.
For the stellar parameters, effective temperatures need to be accurate to within 5-10\%, which is routinely achieved in non-LTE model atmosphere modelling. 
More cumbersome is the estimated $\log g$ that determines the spectroscopic mass.
In Sect.\,\ref{s:multi} we show that in order to be able to derive accurate $\log g$ the UV alone does not suffice, and optical Balmer lines are mandatory (see below). 
Arguably the least well-constrained parameter is the wind mass-loss rate. While UV\,P Cygni lines offer relatively accurate values of the terminal wind velocity, $\sim$10\% \citep{1990ApJ...361..607P}, 
uncertainties in empirical mass-loss rates are about an order of magnitude due to the roles of respectively micro-clumping and macro-clumping \citep{Fullerton:2006,oskinova07,Sundqvist:2018} Clearly such huge uncertainties are not acceptable when building reliable stellar evolution and populations synthesis models. From our experience in stellar modelling, such as the experiments performed in Fig.\,\ref{f:mesa}, we conclude that we need the mass-loss rate to be accurate to 0.3\,dex.

As the mass-loss rate is a multi-variate function of stellar parameters, such as $L$, $M$, and $T_{\rm eff}$, the accuracy requirements on the stellar parameters need to be at least as good as those for the mass-loss rate. Accuracies on $T_{\rm eff}$ are easily within 10\%, though precisions of 1\,kK are sometimes quoted. Similarly, $\log{L}$ precisions of 0.1 dex are feasible. However, the real culprit is the stellar mass $M$ which can be obtained from $\log g$ spectroscopically, but which has a long history of uncertainty, culminating in systematic differences between these spectroscopic masses and evolutionary masses of the order of a factor $\sim$2 (Herrero et al. 1992). In Sect.\,\ref{s:multi} we show that the optical regime is absolutely critical to measure $\log g$.

\begin{figure*}
    \centering
    \includegraphics[width=0.85\textwidth]{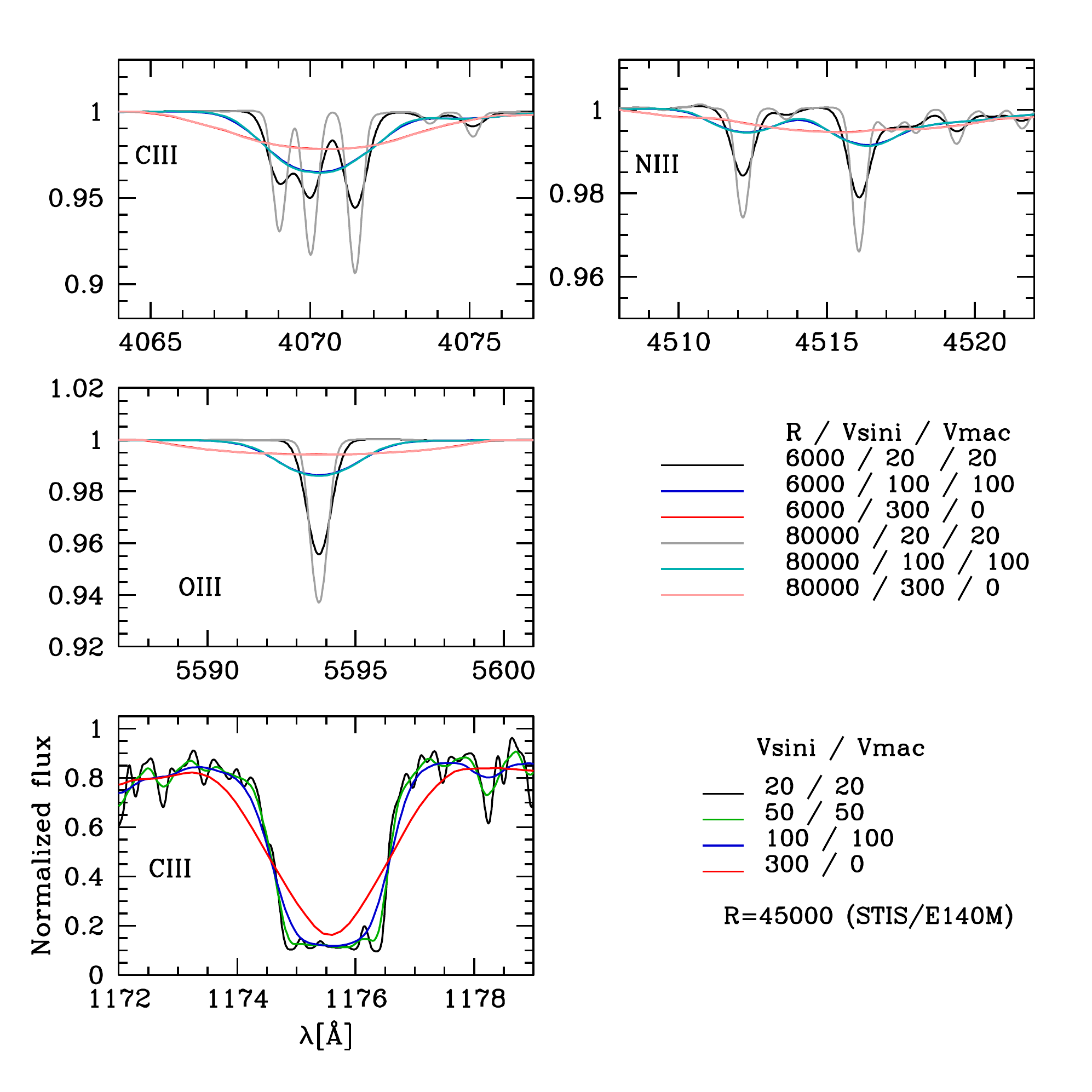}
     \caption{Effect of spectral resolution and rotational velocity on three sets of optical lines classically used to determine C, N, and O abundances. In each panel, the initial {\sc cmfgen} model has T$_{\rm eff} = 31\,000$~K, $\log g=3.6$ and 0.2~$Z_{\odot}$. The model is degraded to a resolution of either 6000 (typical for our X-shooter UVB spectra) or 80\,000 and further convolved with three rotational velocities (20, 100, and 300~km~s$^{-1}$). No additional macro-turbulent broadening is considered. Also plotted is the UV \ion{C}{iii}~1176 line. Here the spectral resolution is that of the {\it STIS E140M\/} grating ($R\sim 45\,000$). 
    }
        \label{f:resol}
\end{figure*}

\section{XShootU data description}
\label{sect::project}

\subsection{Target selection}

The first objective of the XShootU project is to create a homogeneous legacy atlas of similar quality and scope as that of ULLYSES. 
The target sample contains 132 LMC stars, 106 SMC targets, and 6 very-low-$Z$ stars in Sextans~A and NGC~3109 (\citealt{2020RNAAS...4..205R}; Table~\ref{table:targets} this paper).

Most ULLYSES targets are O-type stars (154), but B-type stars (72) and WR/Slash stars (18) are also included. Figure~\ref{fig:skymap} displays the positions of the targets on the sky and Figure~\ref{f:sptype} shows the distribution of spectral types. 
The SMC targets have masses in the range of $10-60~M_{\odot}$, whereas the LMC targets have masses in the range of $15-150~M_{\odot}$.  
A subset of the ULLYSES targets have previously been spectroscopically analysed (see Table \ref{table:params}). These heterogeneous pre-ULLYSES data are presented in the Hertzsprung-Russell (HR) diagram in Figure~\ref{SMC-LMC-HRD} (filled symbols). Estimated parameters for targets lacking contemporary analyses are also shown (open symbols).

XShootU obtained a complementary data-set over the optical to NIR wavelength range for all ULLYSES targets that have not previously been observed with X-shooter. This resulted in a sample of 129\footnote{These numbers are slightly different as a few archival X-Shooter data-sets were available.} LMC stars and 103 SMC stars. In addition, three very low-$Z$ stars were included in the sample. 

The ULLYSES and XShootU data-sets are not taken simultaneously 
in time, although, apart from a few exceptions, 
the vast majority of ULLYSES sources are not known variables. 
In reality, most stars are variable to some level, so care still needs to be taken 
when interpreting the data, but we do not anticipate this to be a massive issue.
Existing ESO Science Archive Facility data are part of both spectroscopic and time-dependent aspects of XShootU. Over half of the ULLYSES targets have no previous high-quality optical spectra. Several have been observed with UVES (15\%) and/or FLAMES ($\sim$50\%), but the wavelength coverage of these FLAMES data is limited. Using (limited) time-sequence data, we will search for binary signatures, and in some cases be capable of disentangling spectra of multi-component systems \citep{2020A&A...634A.118M}. 

\subsection{XShootU observing strategy}

\subsubsection{Wavelength Coverage}

ULLYSES obtained moderate resolution
spectra of OB stars with selected wavelength settings of the {\it COS G130M, COS G160M, STIS E140M, COS G185M,} and {\it STIS E230M\/} gratings in the far- and near-UV during {\it HST\/} cycles $27 - 29$. In order to complement this UV range, similar quality optical/NIR 
spectroscopy was carried out with the X-shooter instrument. This slit-fed (11\arcsec\ slit length) spectrograph provides simultaneous coverage of the wavelength region between $300 - 2500$~nm, divided into three arms; UVB ($300\lesssim \lambda \lesssim 500$~nm), VIS ($500 \lesssim \lambda \lesssim 1000$~nm), and NIR ($1000 \lesssim \lambda \lesssim 2500$~nm). 
X-shooter's wide wavelength coverage made it the instrument of choice for the purpose of building an optical-NIR legacy data-set. 

\subsubsection{Spectral Resolution}

The X-shooter slit widths were chosen to obtain a spectral resolution of $R = 5000-10\,000$, required for estimating the stellar parameters. Each target was observed with a set of $0\rlap{.}^{\prime\prime}8$ (UVB, $R=6700$), $0\rlap{.}^{\prime\prime}7$ (VIS, $R=11\,400$), and $0\rlap{.}^{\prime\prime}6$ (NIR, $R=8100$) slit widths, matching also the average seeing conditions on Paranal. The slit position angle was set by default to parallactic angle, but when necessary a fixed position angle on the sky was used to optimally avoid nearby sources from entering the slits. 

Figure~\ref{f:resol} illustrates that although the spectral resolution is only medium, the determination of surface abundances, including nitrogen (N), carbon (C), and oxygen (O) should be feasible. It is the projected rotational velocity of some of the stars that will limit such studies. The higher \vsini, the broader the lines, which become challenging to identify at very high \vsini. Even with a S/N around 100 (see justification below) most lines would be undetectable at high \vsini. On the other side of the distribution, lines are only partially resolved at low \vsini. Consequently, the determination of accurate low \vsini\ values is not feasible and additional higher spectral resolution data are needed for this subset (see Appendix WG\,11) description of auxiliary Magellan/MIKE data), although the combined analysis of optical and UV lines can partially alleviate this limitation. The bottom panel of Figure~\ref{f:resol} illustrates the effect of rotational broadening of \ion{C}{iii}~1176, a line complex relevant for the determination of \vsini\ (e.g., \citealt{2013A&A...555A...1B}). Above 100 km~s$^{-1}$, the components of the multiplet are blended, while at lower rotational velocities they are resolved individually.

\begin{figure*}
    \centering
    \includegraphics[width=0.8\textwidth]{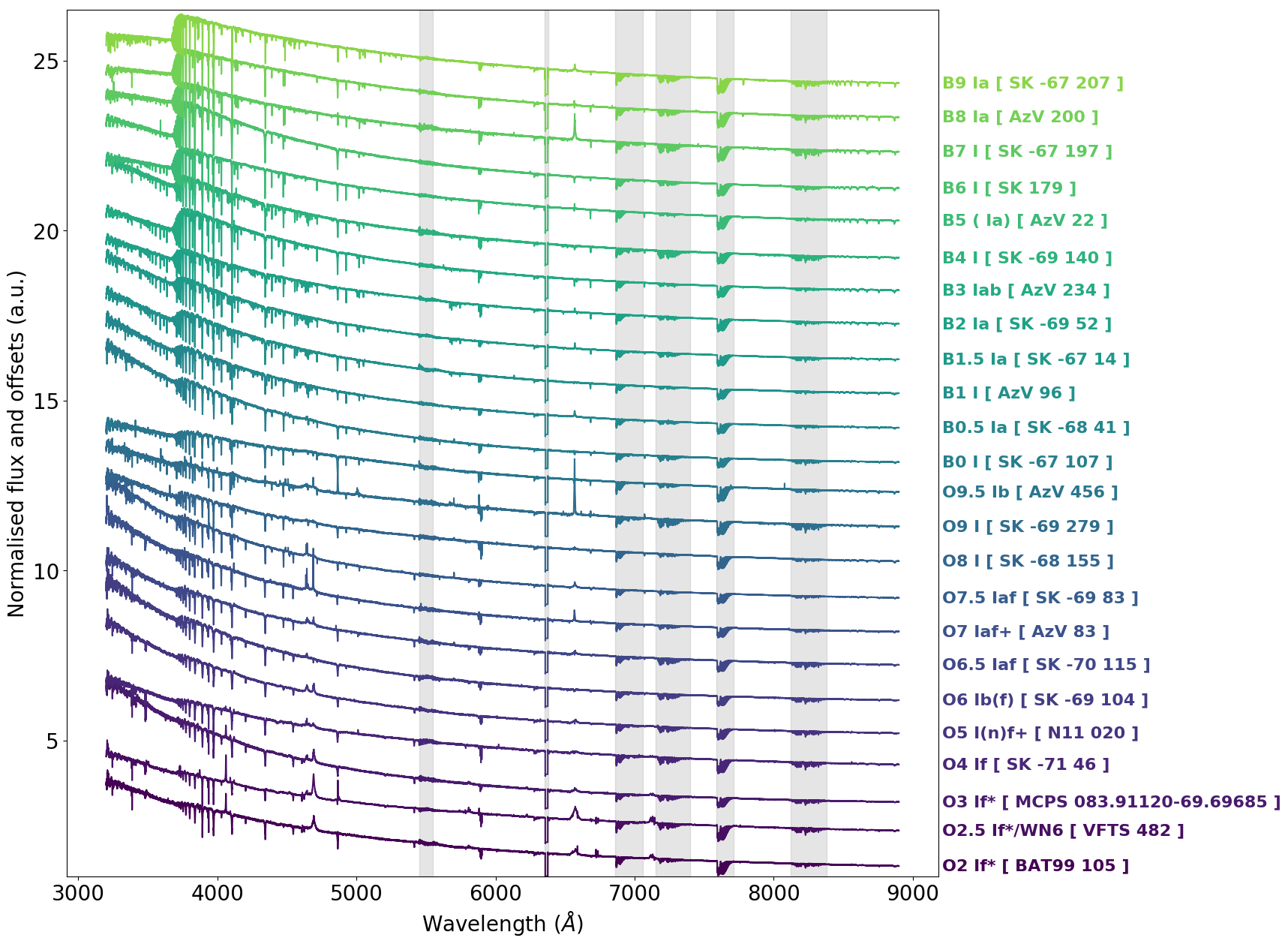}
    \includegraphics[width=0.8\textwidth]{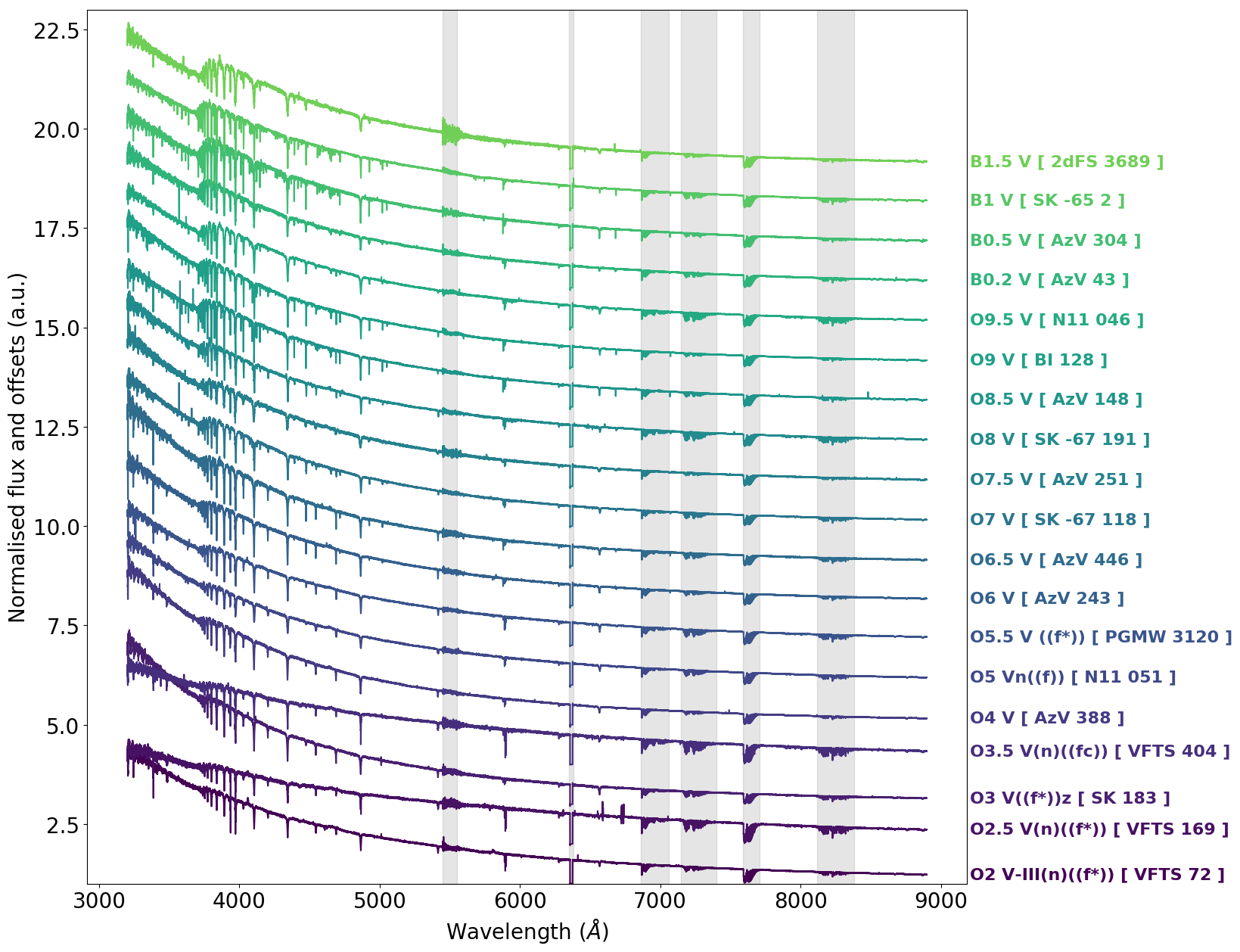}
    \caption{Reduced X-shooter spectra for a range of spectral types of single-star supergiants (top) and dwarfs (bottom). For illustration purposes the flux of each spectrum was divided by its mean value and an arbitrary offset was added. The grey regions correspond to the UVB-VIS common wavelength coverage ($\sim5500~\AA$), a gap due to bad pixel masking ($\sim6360\,\AA$), and telluric absorption. Minor manual treatment to remove strong cosmic rays was performed.}
    \label{f:sample_spectra}
\end{figure*}

\begin{figure*}
    \centering
    \includegraphics[width=0.94\textwidth]{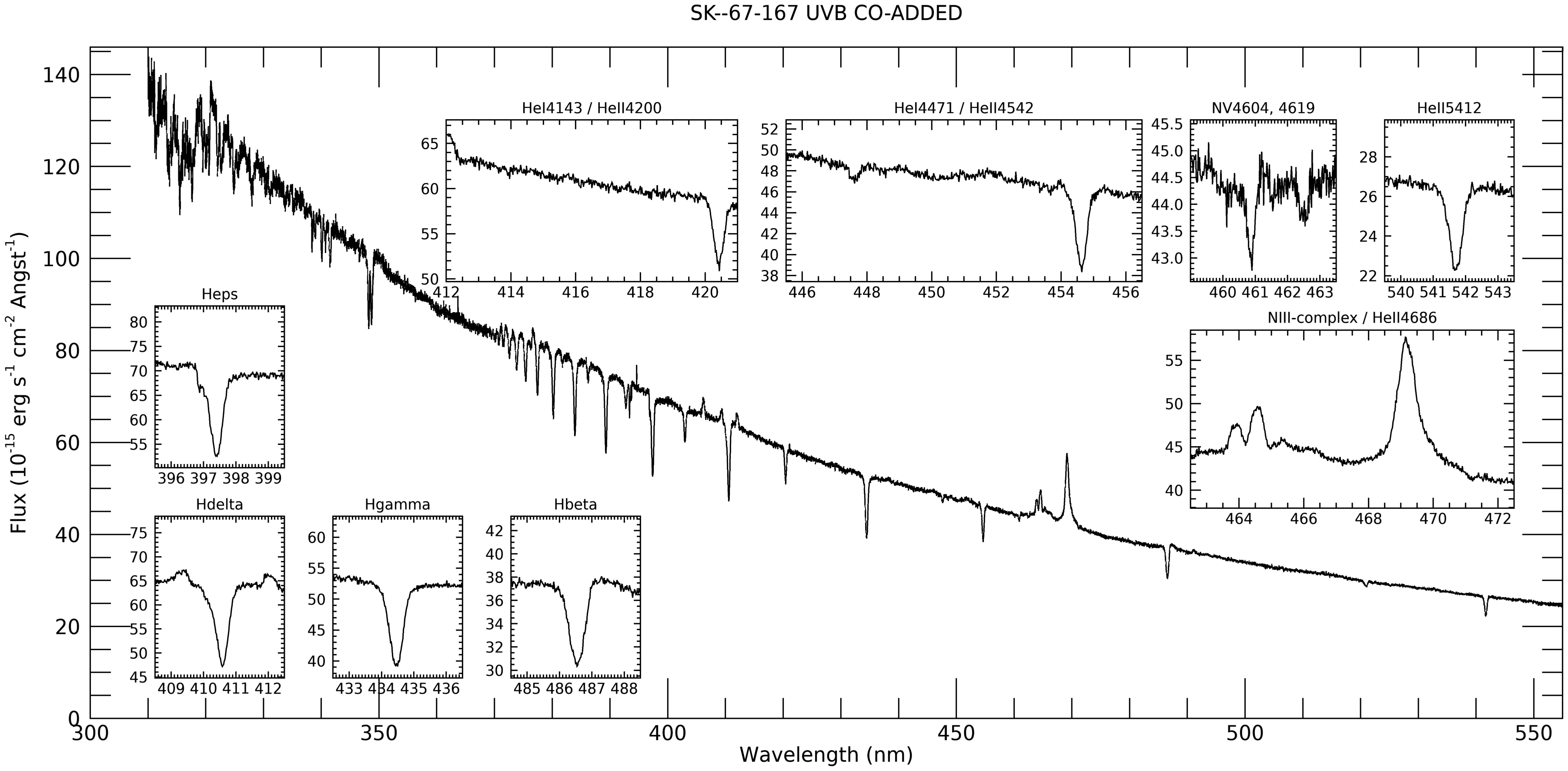}
    \caption{Xshooter/UBV spectrum of Sk $--67^{\circ} 167$ (O4\,Inf$+$) in the LMC, 
    including zoom-in views of key spectroscopic diagnostics.}
        \label{f:QC}
\end{figure*}

\subsubsection{Signal-to-Noise Ratio}

Some VLT instruments provide higher spectral resolution in certain wavelength regimes  (e.g., UVES), but to build a homogeneous database with a wide spectral coverage could only be achieved with X-shooter. 
In addition to the wide wavelength coverage, a high signal-to-noise ratio (S/N) in addition to sufficient spectral resolution are essential to determine the fundamental stellar parameters and the abundances for various temperature regimes populated by OB and WR stars.
For the preparation of the X-Shooter proposal, we estimated the required S/N, experimenting on a typical 
SMC mid-O dwarf/giant with {\sc cmfgen} model spectra degraded to X-shooter's spectral resolution.
We found that the determination of basic stellar parameters such as $T_{\rm eff}$ and $\log g$ became prohibitive for quantitative interpretation if the S/N drops below $100$ per resolution element. 
To ensure maximal scientific return of XShootU, we achieved a S/N of $>100$ in the continuum in the UVB and VIS for all MC targets. 

\subsection{Data Reduction}

A detailed description of the data reduction is provided in a paper associated with Data Release 1 (DR1; Sana et al.\ 2023, XShootU {\sc II}). Here, we provide a brief summary, focusing on the UVB and VIS spectra. The data reduction of the NIR spectra requires additional efforts and will become part of DR2.

The initial data reduction was performed using the ESO X-shooter pipeline v3.5.0 \citep{2011AN....332..227G}. The pipeline carried out the standard steps of bias, flat, wavelength calibration, spectral rectification, cosmic ray removal, sky subtraction, flux calibration, and extraction of a 1D spectrum.
The wavelength calibration was performed using a physical model, i.e., the transformation from pixel to lambda space was optimized through the analysis of a multi-pinhole ThAr (UVB, VIS) or penray (NIR) lamp frame. The predicted positions of the lines were fitted using a 2D Gaussian to recover the actual positions on the frame. 

The pipeline-reduced data were subsequently flux calibrated using a set of 6 standard stars observed during the same or adjacent nights. We found that the stellar models used by the public pipeline \citep{Moehler2014} resulted in small (on the order of a few percent) changes in the Balmer line profiles depending on the standard star. This potentially impedes accurate $\log g$ measurements. In addition, the spectral energy distribution for some of the standard stars could be optimized. We decided to use new stellar models and new fit points to derive the response, starting with models used by {\it HST\/} for their fundamental flux standards GD~71 and GD~153 \citep{Bohlin2020}.\footnote{The {\it HST\/} models are available at \url{https://www.stsci.edu/hst/instrumentation/reference-data-for-calibration-and-tools/astronomical-catalogs/calspec}.} We then reduced observations of the other 5 standard stars taken between October 2020 and April 2021 as if they were science objects, with the response determined by close-in-time observations of GD~71. Those spectra were then co-added to create high S/N spectra that were used to derive improved stellar models.
The XShootU spectra were obtained with narrow slits. 
To obtain absolute flux calibrated spectra, corrections were applied for slit losses due to seeing and image quality across the detector and by re-scaling to existing photometry.
The achieved accuracy is typically better than 5\%. 

Telluric correction was performed using the molecfit tool v3.0.3 \citep{2015A&A...576A..77S,2015A&A...576A..78K} for the VIS arm and generally leads to good results. For the MCs targets, we fitted the atmospheric model directly to the science spectra, as the S/N on the continuum is high enough to ensure a better correction than using a telluric standard star to compute the model. 
The regions with very deep O$_2$ telluric absorption lines at $\sim$760 nm and sometimes the one at $\sim$690 nm are poorly corrected and the correction of the H$_2$O bands at $\sim$950 nm leaves strong residuals. The correction from telluric lines around the [OI]~6300\AA\ line is always good. 

Results of the data reduction are shown in 
Figure~\ref{f:sample_spectra}. Here a sample of reduced X-shooter spectra is presented to highlight a sequence from the earliest to the latest spectral types for supergiant and dwarf targets. The spectra shown are single-epoch in order to avoid confusion in co-added spectra due to potential variability. 
Telluric correction (grey regions) and proper cosmic-ray removal was not performed for this plot, but will become part of the first data release (Sana et al. 2023, XShootU {\sc II}). 
An example of an O4 supergiant spectrum is shown in Fig.\,\ref{f:QC} on a improved scale, focused on wavelength regions in which telluric corrections are not needed. 

\section{Multi-wavelength analyses}
\label{s:multi}

\begin{figure*}
    \centering
    \includegraphics[width=0.95\textwidth]{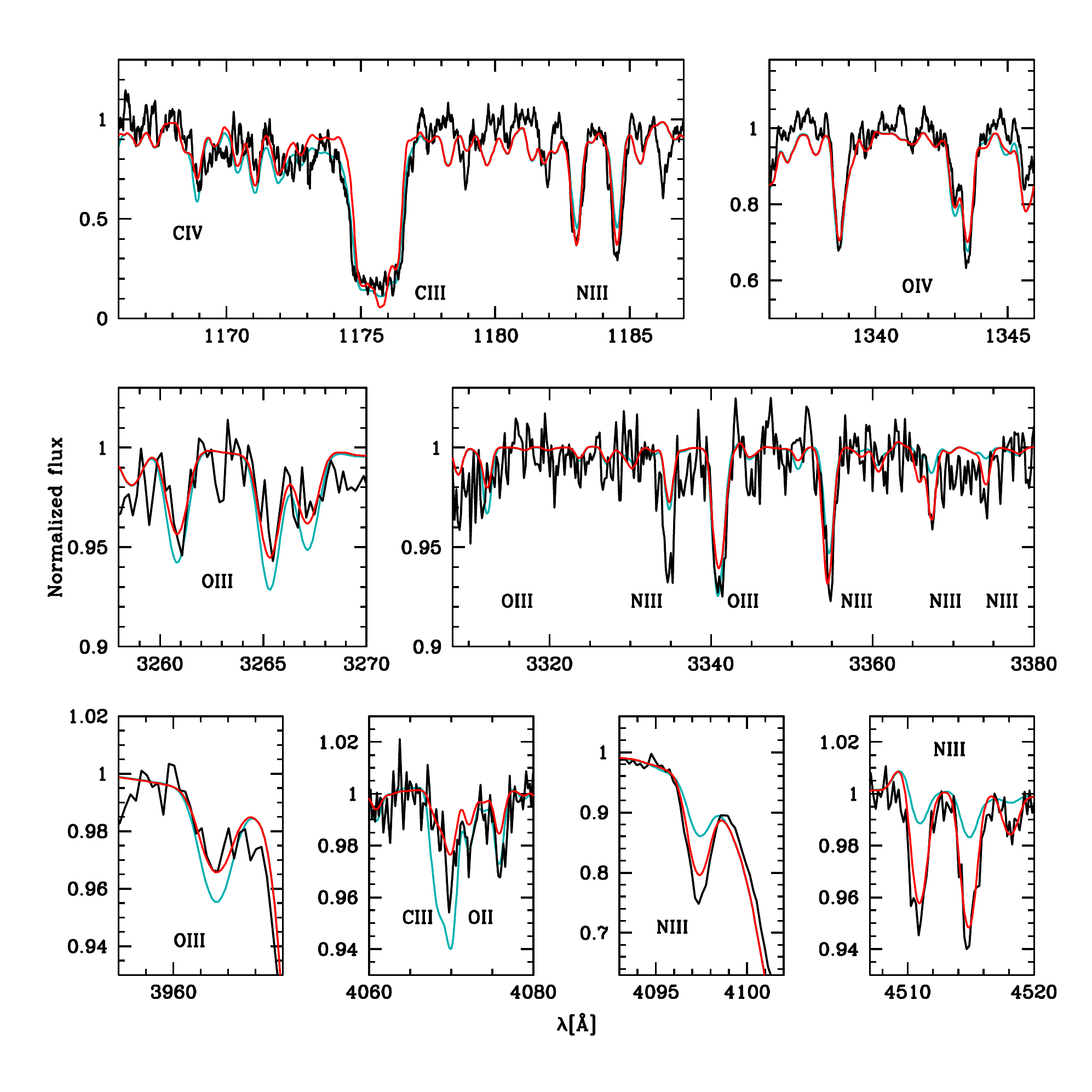}
    \caption{Comparison between the observed HST (top) and X-shooter spectrum (black line) of selected lines of C, N and O lines for AzV\,327 (O9.5 II-Ibw) with two models (colored lines). The light blue line is for solar-scaled abundances (factor 1/5) while the red model has the following scaling: C abundance decreased by a factor 3.8, N abundance increased by a factor 4.5, and the O abundance decreased by a factor 1.6. 
    The models were computed with the non-LTE CMFGEN \citep{1998ApJ...496..407H} atmosphere code.}
    \label{f:ex_cno}
\end{figure*}

In Sect.\,\ref{s:req} we showed that the low-$Z$ environment of the SMC can be considered rather 
characteristic of the Early Universe, with low mass-loss rates, and the potential for rapid rotation and bluewards evolution, while the LMC sample is more characteristic of today's Universe, with higher mass-loss rates, slower rotation, and classical redwards stellar evolution. In reality, the situation is more complex, as the mass-loss rates is a function of $\dot{M} = f(Z,L,M,T_{\rm eff}, v\sin{i})$, which implies we need to obtain stellar and wind parameters over a large parameter space, including not only the O-star regime, but also the B supergiant regime, where the bi-stability jump may increase mass-loss rates (Vink et al. 1999), or not (Bj\"orklund et al. 2021).
Moreover, stellar evolution models depend on interior mixing, and stellar abundances can be utilized to test the efficiency of (rotational) mixing. 

\subsection{Diagnostics in the UV and Optical range}

\begin{figure*}
    \centering
    \includegraphics[angle=270,width=0.95\textwidth]{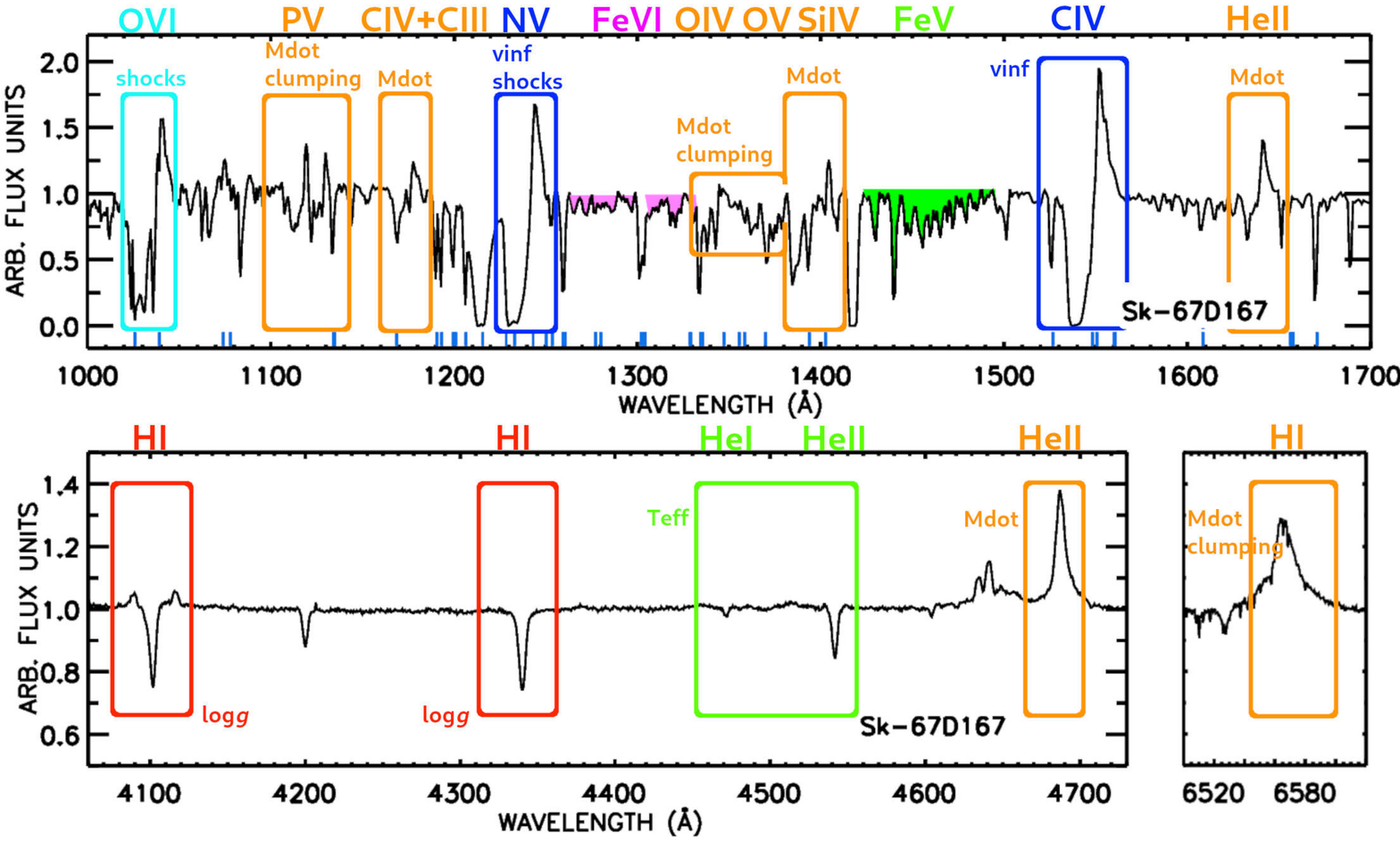}
    \caption{UV (top) and optical (bottom) spectrum of the LMC star Sk-67~167 (O4~Inf+). The UV spectrum consists of {\it STIS E140M\/} observations taken as part of the ULLYSES project and archival {\it FUSE\/} data. The optical spectrum was obtained with X-shooter. A selection of diagnostics for stellar and wind parameters are highlighted. Note that these diagnostics can vary with spectral type. The ticks at the bottom of the UV spectra mark the position of interstellar lines.}
    \label{f:uv-opt}
\end{figure*}

To start with the latter, He/H abundances can only be determined from the optical since H/He lines in the UV are dominated by strong interstellar features (e.g. Lyman alpha) so the Pickering-Balmer lines in the optical are critical for the He/H ratio. 

Abundances of C, N, and O can in principle be determined from the UV range only (e.g. Bouret et al. 2003, 2013) but most lines are also sensitive to winds, especially as one moves away from the main sequence. Fig.\,\ref{f:ex_cno} shows an example where winds are sufficiently weak for such a determination. The plot also highlights that the optical range contains more lines from these elements, and these lines depend far less on wind properties than those in the UV. Using more lines reduces the systematic uncertainties in the determinations. 
Figure\,\ref{f:ex_cno} highlights that it is still challenging to obtain a perfect fit for all lines of the same element, but the availability of more lines helps identifying potential shortcomings in the atmosphere models. It also allows for a better determination of errors associated with abundance determinations, which is crucial to interpret stellar evolution predictions of interior mixing.
A full error determination will follow in a dedicated paper, but we could already say that typical error bars are 15-30\%, sometimes up to 50\%, with these two data-sets combined (see Bouret et al. 2021), easily satisfying our science requirements.

Another key aspect of combining the ULLYSES and XShootU data-sets is that 
it allows stellar and wind parameters to be derived self-consistently using both optical and UV diagnostics, which was lacking in surveys such as VFTS.
For O-type stars, there are no conclusive diagnostics in the UV to derive effective temperatures and gravities. The wind profiles of O\,{\sc iv}, O\,{\sc v}, and N\,{\sc iv} impose a minimum effective temperature, but they are also sensitive to the mass-loss rate and clumping properties. Figure\,\ref{f:uv-opt} highlights which parameters can be determined from specific UV as well as optical lines for an LMC O supergiant. 

\begin{figure*}
    \centering
    \includegraphics[width=0.95\textwidth]{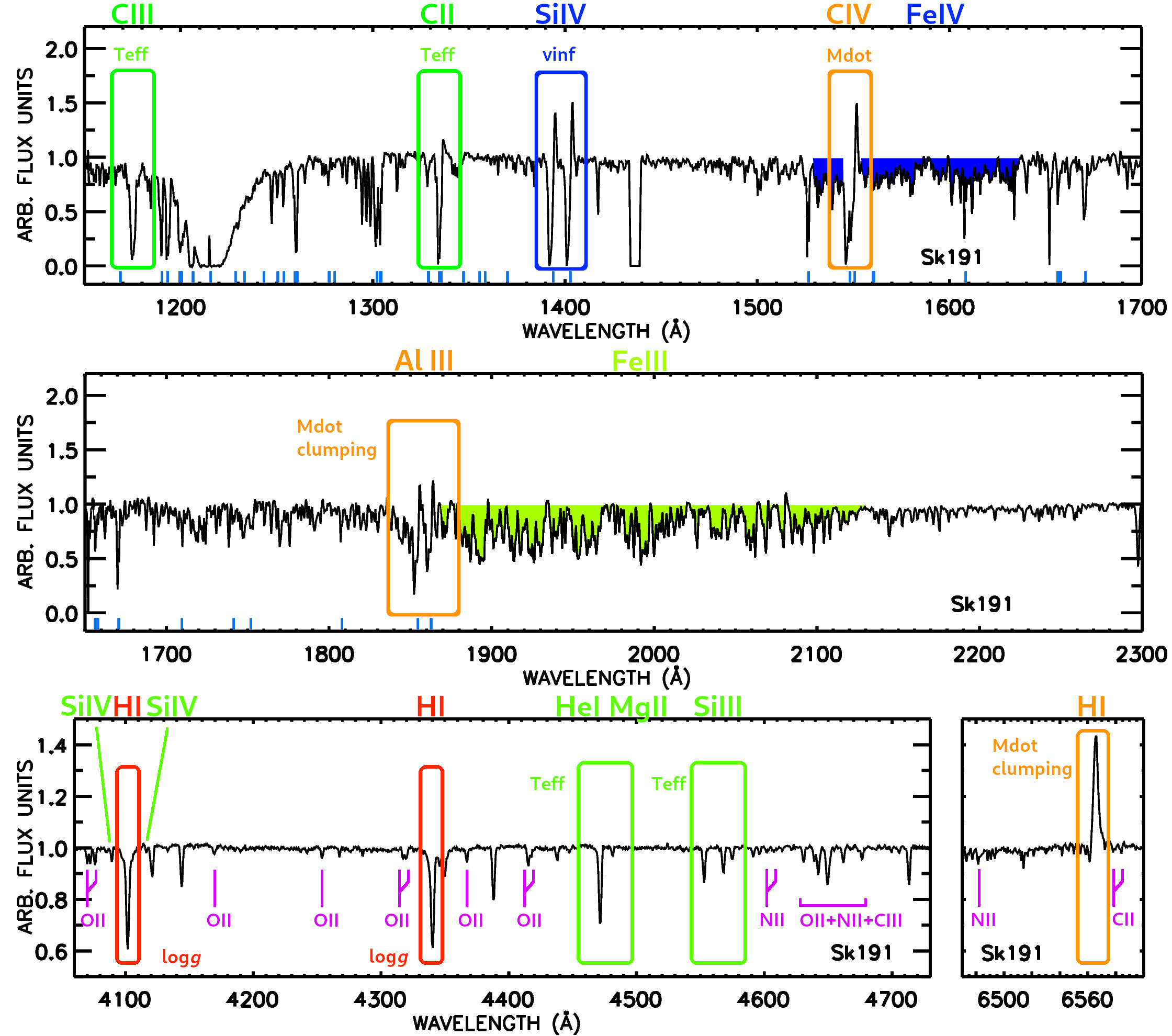}
    \caption{UV (top, middle) and optical (bottom) spectrum of the SMC star Sk191 (B1.5 Ia). The UV spectrum consists of FUSE, STIS E140M and STIS E230M observations compiled as part of the ULLYSES project. The optical spectrum was obtained with X-shooter. Similarly to Fig.\,\ref{f:uv-opt} interstellar transitions and a selection of stellar and wind diagnostics are highlighted. In addition, metallic lines that can be used to measure abundances are marked in purple.}
        \label{f:uv-opt-B}
\end{figure*}

As 30\% of the sample involves B stars rather than O-type stars, we also show Fig.\,\ref{f:uv-opt-B} highlighting which parameters can be 
determined from the UV versus optical part for B supergiants \citep{Crow06,Firn12,McEvoy15}. The combination of UV and optical spectra is even more powerful to constrain the physical properties of B-type stars. As in the case of O-stars their UV spectrum alone does not contain diagnostics for gravity. 
Effective temperature could to first order be constrained by comparing the C{\sc ii}/C{\sc iii} lines and the Fe{\sc iii}/Fe{\sc iv} line forests, although the C lines are also sensitive to mass loss, and Fe transitions depend also on $\log g$. 
The optical range offers cleaner $T_{\rm eff}$ diagnostics from the ionization balance of Si{\sc ii}/Si{\sc iii}/Si{\sc iv} (secondarily, the comparison of He{\sc i} and Mg{\sc ii}) and gravity (e.g. H$\gamma$ and the higher Balmer lines). 
The numerous metallic lines in the optical can be used to determine abundances and in the case of the strongest transitions (e.g. Si{\sc iii}) micro-turbulence, and projected rotational velocity. 
The joint UV $+$ optical range offers several mass-loss rate and clumping diagnostics, with the Si{\sc iv} doublet being the best wind velocity indicator for early B supergiants.

\begin{figure}
    \includegraphics[width=0.98\columnwidth]{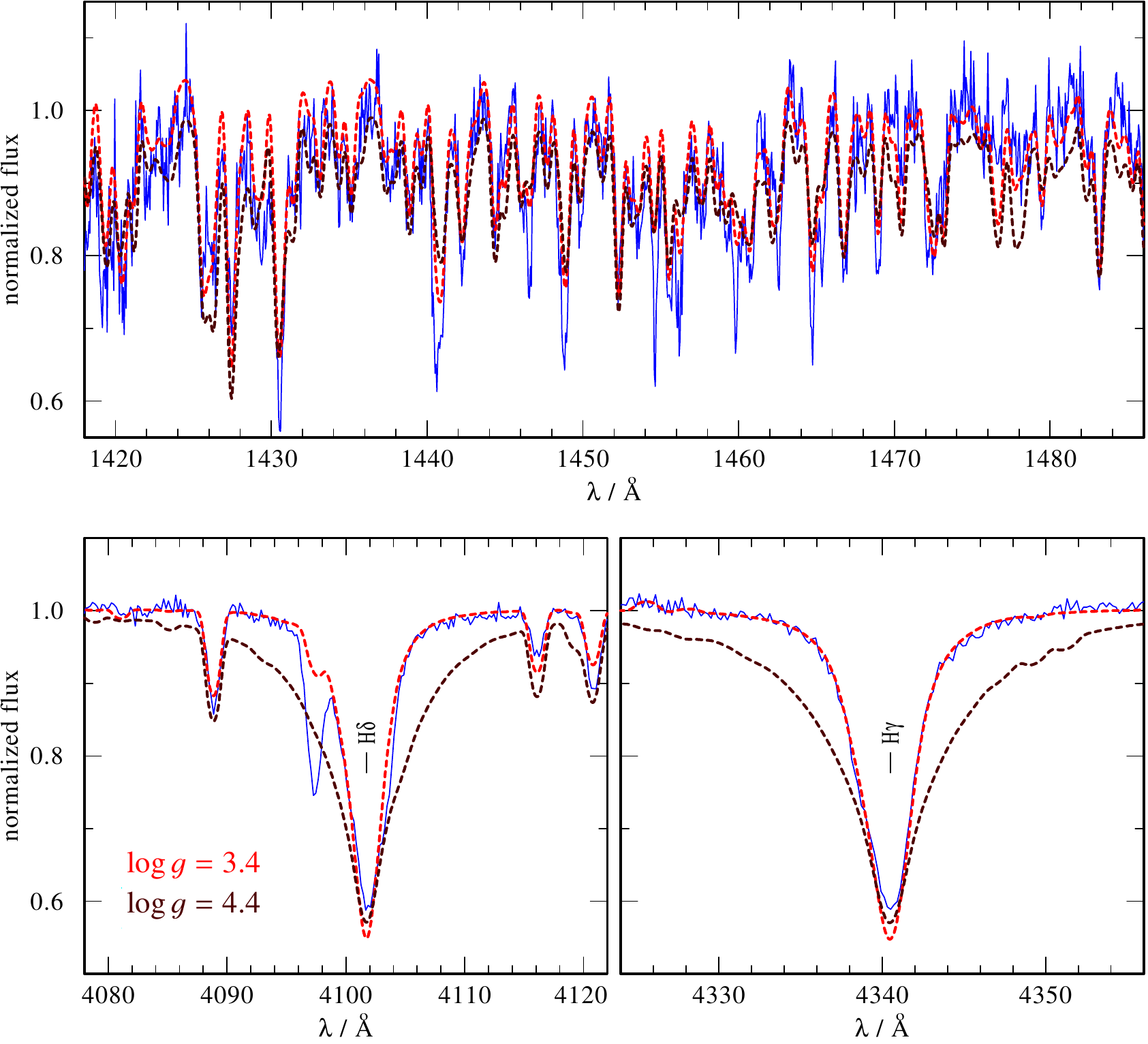}
    \caption{O giant (AV186, O8.5III). The best fitting model is for $T = 33$ kK and $\log g=3.4$. 
    Note that while a ten times higher mass star with $\log g = 4.4$ would be indistinguishable in the UV, it would completely fail to reproduce the optical Balmer wings. The model was computed with the PoWR \citep{Sander:2017} non-LTE code.} 
    \label{f:uvopt}
\end{figure}

In order to further quantify the need for optical X-Shooter spectra in spectroscopic analyses, 
we present an example analysis for an O8\,III SMC giant in Figure~\ref{f:uvopt}. 
The figure shows both the UV part of the spectrum and some Hydrogen Balmer lines that are routinely utilized to derive $\log g$ values. It can easily be seen
that while high and low $\log g$ model values reproduce the UV spectra equally well, the
optical is critical for accurate $\log g$ determination.
While the complexity of the spectroscopic analysis is beyond the scope of this paper, the key point is that 
UV-only fits yield very poorly constrained surface gravities, which result in enormous uncertainties 
on spectroscopic masses.

In addition to the uncertainty in surface gravity, it is also appropriate to mention that for stars with strong winds, such as supergiants, the H$\alpha$ Balmer line is a key mass-loss and clumping factor diagnostic. When only accounting for micro-clumping and the UV part of the spectrum, \cite{Fullerton:2006} showed that clumping factors were uncertain by factors of up to a hundred, and mass-loss rate reductions could easily be an order of magnitude. Only when accounting for the optical H$\alpha$ line and macro-clumping (see Appendix WG\,4), 
\cite{oskinova07} showed that mass-loss rate uncertainties were significantly smaller, by a factor of 2 or so, and from the additional optical H$\alpha$ line, clumping factors are usually estimated to be lower of order 6-8 (e.g. Ramirez-Agudelo et al. 2017).

\subsection{Spectroscopic analyses tools and procedures}

XShootU is coordinating spectral modelling efforts for massive stars on a world-wide scale never witnessed before in the massive-star community. 
Before we can scale-up the analysis to hundreds of massive stars with hugely varying spectral and wind properties over the entire hot part of the HR diagram, it is paramount that codes and analysis techniques are tested and compared as a function of stellar parameters and metallicity. 

The spectral analysis of massive stars is rather intricate due to the highly non-local thermodynamic equilibrium (NLTE) conditions in their turbulent, supersonic atmospheres.
Over the past decades, a number of highly complex, yet successful, model atmosphere codes have been developed, e.g., CMFGEN \citep{1998ApJ...496..407H}, PoWR \citep{ 2003A&A...410..993H,2015A&A...577A..13S} and FASTWIND \citep{1997A&A...323..488S,2005A&A...435..669P}. Although these codes have previously been applied to various sets of observations, only more recently have they been used for larger samples (e.g., \citealt{2017A&A...600A..81R,2017A&A...601A..79S}) due to the efficiency of numerical methods (applying certain physical approximations), and efficient spectral automated analysis tools including genetic algorithms and grid-based $\chi^2$ approaches. 

At virtual and on-site Lorentz workshops in 2021 and 2022 (and various additional virtual XShootU meetings) preliminary comparisons of analyses with the various non-LTE codes were performed by modelling subsets of O-stars. Agreement was reached on a common methodology to perform the spectroscopic analyses within the XShootU Project.

This recommended procedure may be summarised as follows: 
 \begin{enumerate}
	\item Use the same reference value for the distance moduli (DM) of the LMC and the SMC. We adopted DM(SMC) = 18.98 mag \citep{2020ApJ...904...13G} and DM(LMC) = 18.48 mag \citep{2019Natur.567..200P}.
	\item Adopt the same photometry ($U, B, V, J, H, K_{\rm S}$, as minimum) for each star (Table\,\ref{table:targets}). Optical photometry is from a variety of sources, whereas near-IR photometry is from VISTA VMC (JKs) \citep{Cioni11}, 2MASS (JHKs) \citep{Cutri03}, or 2MASS 6X (JHKs) \citep{cutri12}. H-band photometry is omitted if JKs values are discrepant between 2MASS and VMC owint to photometric variability or crowding.
 
	\item Adopt the same literature source for the bolometric correction. The relation for bolometric correction as a function of \teff\ and metallicity from \cite{2003ApJS..146..417L} is adopted for now. This relation may be updated in the course of this project.
    \item Adopt the same reddening approach. Key references for the Milky Way foreground are \cite{Fitzpatrick19}, over earlier works by \cite{Seaton79,Cardelli89}, though Galactic foreground extinction is modest towards the MCs.
    
For the MC contributions, \cite{Gordon03} is preferred, recognising earlier contributions from e.g. \cite{Howarth83,Fitzpatrick86} to UV laws in the LMC, and for the SMC e.g. \cite{Prevot84,Bouchet85} to the UV and optical/IR respectively. For the FUSE range, \cite{Gordon09} and \cite{Cartledge05} are recommended for the Milky Way, and MCs, respectively.
	\item Adopt the same baseline LMC and SMC abundances. 
	Several abundance ratios in the MCs are notoriously non-solar, which is especially true for \mbox{CNO} in the SMC, and the use of scaled-solar values should be avoided when possible. 
    Thus, the adopted abundance values were derived from an average of different determinations, e.g., from stars, \ion{H}{ii} regions, and SN remnants.
    The recommended values are listed in Tables \ref{LMCabundances} and \ref{SMCabundances}.
    For several species, however, we need to agree on default abundances, given the absence of lines in spectral ranges covered by ULLYSES $+$ XShootU. 
    In such cases, scaled-solar values need to be adopted, using 0.5\,$Z_{\odot}$ and 0.2\,$Z_{\odot}$ scaling factors for the LMC and SMC, respectively \citep{2009ARA&A..47..481A}. 
	\item Whenever possible, adopt the same description of macro-turbulence for line broadening. The recommendation is to adopt a radial-tangential description of macro-turbulence \citep[e.g.,][]{2014A&A...562A.135S}. 
	\item Adopt similar wind clumping implementation. We agreed to use the same parametric description as implemented in CMFGEN \citep{1998ApJ...496..407H} for the standard derivation of the mass-loss rate. 
 Clumping is predominately treated in the ``microclumping'' approximation, assuming a void interclump medium. 
 The volume filling factor $f_\text{V}$ has a velocity-dependent behavior that goes as:
\begin{equation}
\label{eq:cmfgen-clumping}
    f_\text{V}(r) = f_{\text{V},\infty} + (1 - f_\infty)\,e^{-\varv(r)/\varv_\text{cl}}
\end{equation}
where $f_{\text{V},\infty}$ denotes the value at $r \rightarrow \infty$. For a void inter-clump medium, the corresponding clumping factor $D \equiv f_\text{cl}$ is simply the inverse value, implying that $D_\infty = f_{\text{V},\infty}^{-1}$. 
The free parameter $\varv_{cl}$ is a characteristic velocity varied to describe the clumping onset. 
 More sophisticated descriptions of the properties and nature of clumping can be implemented in the codes \citep[e.g.,][]{oskinova07,Hawcroft:2021,2021MNRAS.504..311F}. These predominately fall under the framework of WG\,4 that focuses on wind structure (see below). 
\end{enumerate}	

The next step is to benchmark the accuracy of stellar parameters derived with the different approaches. For this, a small set of stars was modelled with various non-LTE codes. The obtained stellar parameters were compared, allowing us to assess to what extent the physical interpretation depends on the modelling tools applied. 
In parallel, we also considered bench-marking of the non-LTE wind codes, i.e. against one another or against mock data, to perform a direct comparison of synthetic spectra computed for the same input model parameters. Alternatively, we could use a model obtained with one code and fit this model with the other codes. 
These various "bench-marking" approaches will provide relevant information of systematic differences between codes, analyses tools, and other differences in approach, which will be detailed in a future (benchmarking) paper (Sander et al. 2023; XShootU IV).

Determination of the wind terminal velocities ($v_{\infty}$) is also taken on.  
As a global wind parameter, this is indeed an essential input for the models of stellar atmospheres used. 
The ULLYSES (UV) data are of particular importance for this task as they are rich in resonance transitions of ionised species which are prime $v_{\infty}$ diagnostics. 
For stars presenting saturated resonance-line UV profiles, considering that the spectral lines remain optically thick at the distances where the wind reaches its maximum speed, \vinf\ can be measured directly by measuring the maximum velocity shift of the absorption component of the UV \ion{C}{iv} resonance doublet \citep[see, e.g.,][]{1990ApJ...361..607P, 1998MNRAS.300..828P}.
Alternatively, the wind speeds can also be measured by fitting synthetic spectra produced using the Sobolev with Exact Integration (SEI) method \citep{1987ApJ...314..726L}. This latest approach is particularly relevant for stars without saturation
in their UV resonance-line profile (although it can also be used in the first case mentioned above). As this paper is being written, a significant fraction of the LMC and SMC star sample has been studied with either method. Results concerning the dependence of \vinf\ with the ambient metallicity or stellar parameters will be presented in dedicated papers (e.g. Hawcroft et al., XShootU III).  

\section{Final Perspectives}

The XShootU project is expected to provide many pieces of data, models, as well as new physics of massive stars in low $Z$ environments. It is important to stress that the overarching aim is to provide a high-quality homogeneous optical database that is complementary to ULLYSES. These legacy data-sets are critical for a correct interpretation of unresolved high $Z$ observations with JWST/NIRSPEC \citep{Curti23,Carnall23,DeBarros19}. 
The next goal is to provide uniformly determined stellar and wind parameters from the combined UV and optical data-sets. 
For this part of the project, it is not only critical to include the correct non-LTE physics, but also to test the various spectral synthesis 
codes and analyses. 

A key science aim is to quantify {\it how} the mass-loss rate declines with decreasing metallicity. This does not necessarily simply imply a determination of a power-law exponent, as the slope may easily vary with stellar parameters or $Z$ itself \citep{2001A&A...369..574V,Kudr02,SV20,Marcolino22,Rickard22}.
In order to obtain an empirical $\dot{M} = f(Z)$ relationship, 
we not only require mass-loss rates and clumping properties, but simultaneously also need to obtain the underlying {\it stellar} parameters ($T_{\rm eff}$, $\log g$, $L$, $M$) as these parameters enable mass-loss properties from a given object in one galaxy to be compared to those from an entirely different object located elsewhere. 

The XShootU data-set, coupled with surveys of \ion{H}{ii} regions in the LMC and SMC such as SDSS LVM\footnote{https://www.sdss.org/dr15/future/lvm/}, represents a significant opportunity to advance the state-of-the-art in our understanding of how stars shape their environment.
LVM is an optical ($3600-10000$\,\AA) integral-field spectroscopic survey ($R\sim4000$) of the Milky Way and the Local Group (LMC/SMC/M31/M33). It will be the first IFU survey to isolate and resolve distinct environments within galaxies, and to cover significant portions of the night sky. New populations synthesis models -- informed by the new physics obtained in the XShootU collaboration -- can be used to remove the stellar contribution from LVM observations of star-forming regions, which will enable studies of the ionized-gas alone (H\,{\sc ii} regions, diffuse ionized gas), over the large dynamic range in size of LVM, i.e., from clusters and clouds (10-50 pc) to the kpc scales of spiral arms, galactic in/outflows, and disk dynamics.

One should be aware that different communities refer to $Z$ in different ways. 
Extra-galactic communities usually work on the basis of nebular oxygen (O) lines, while stellar astronomers are sometimes able to derive Fe abundances of individual stars. It is predominately the Fe abundance that sets the mass-loss rate, while intermediate mass elements such as O set the wind terminal velocity \citep{Vink99,Puls2000}. 
While the [$\alpha$/Fe] ratio in Local dwarf galaxies such as IC\,1613 is generally found to be sub-solar \citep{Taut07,garcia14}, \cite{Steidel16} and \cite{Strom22} showed O to be {\it enhanced} in comparison to Fe for galaxies at intermediate redshift ("Cosmic Noon"). 
This is thought to be due to the time delay in the production of Fe from Type Ia supernovae with respect to $\alpha$ elements released by massive stars.
It may therefore become relevant to consider more detailed abundance patterns than simply scaling all metals with the solar-abundance pattern, i.e. to make a clearer distinction between [Fe/H] and $\alpha$ elements for $\dot{M}$ calibrations of massive stars, as indicated in Table 5 in Vink et al. (2001). The ULLYSES and XSHootU sample will uniquely provide the opportunity to investigate potential differences between [Fe/H] and $\alpha$ elements with non-solar patterns.

When we are eventually able to provide a reliable empirical $\dot{M} = f(Z)$ relation we can compare these findings to theoretical predictions, and this will inform us on how to treat mass loss more reliably in models of stellar evolution -- and thereby also in feedback and population synthesis studies -- at low $Z$. 
It is currently unclear if most of the mass loss takes place in $Z$-dependent winds, or in $Z$-independent eruptions or binary interactions. 
Despite the possibility of significant $Z$-independent mass loss, there is 
ample evidence that massive-star evolution in metal-poor environments proceeds very differently from our Milky Way. 
\citet{2012MNRAS.421.1043S} and \citet{2015ApJ...801L..28K} found strong nebular He\,{\sc ii} emission lines in low-$Z$ galaxies, but not in metal-rich ones.
Long-duration gamma-ray bursts (GRBs; e.g., \citealt{2004A&A...419..927V}),
superluminous SNe (e.g., \citealt{2011Natur.474..487Q,2012Sci...337..927G,2015MNRAS.452.1567C}),
and broad-line Type\,Ic SNe occur preferentially in low-$Z$ dwarf galaxies (e.g., \citealt{2019A&A...623A..26P}),whereas ordinary Type\,Ic SNe avoid dwarf galaxies.

The spectacular merger of two $\sim30~M_{\odot}$ black holes observed through the detection of GWs by the LIGO observatory \citep{2016PhRvL.116x1102A} most likely originated from a system that had an initial $Z$ similar to that of the SMC \citep{2016ApJ...818L..22A}. This conclusion is inextricably linked to the predicted reduced stellar wind mass loss at low $Z$ (e.g., \citealt{2001A&A...369..574V}).
Proposed channels that may have lead to GW\,150914 (and other more recent events) involve mass transfer and common envelope evolution (e.g., \citealt{2016Natur.534..512B}) and ``chemically homogeneous evolving'' (CHE) systems \citep{2016MNRAS.458.2634M}. The latter model is linked to rapid spin rates, leading to rotationally-induced mixing of the stellar interior \citep{1987A&A...178..159M,2005A&A...443..643Y}.
Predictions of single star evolution have shown this process to be increasingly important at lower $Z$ \citep{2011A&A...530A.115B} due to lower mass-loss rates, and correspondingly less angular momentum loss. Support for the existence of CHE can be found in the properties of a small fraction of WR stars \citep{2009A&A...495..257M,2015A&A...581A..21H}, but CHE still lacks unambiguous observational confirmation in the O-star regime (see for example \citealt{2004ApJ...608.1028W, 2019ApJ...880..115A, 2021A&A...651A..96A} and \citealt{2013A&A...555A...1B} for candidates). 

The properties of GW events and exotic SNe illustrate the key physics involved: spin rates and rotationally-induced mixing, CHE, and wind mass-loss rates. 
The XShootU sample consists of uniformly estimated spectral parameters, of objects previously classified in the literature to varying extents. 
Further analysis will include, for instance a determination of an upper limit to the number of chemically-homogeneous evolving (CHE) stars for which three basic observables need to be fulfilled: (i) a  peculiar HR diagram location, as chemically homogeneous stars evolve {\it bluewards} instead of redwards (ii) higher average \vsini, as they are thought to be caused by rapid rotation, and (iii) special chemical-abundance patterns showcasing chemical mixing.  

These are just some of the studies that are being prepared and their results will be published in research articles of the XShootU series. It should be emphasised that some of the analysis is already ongoing, but there is ample space for new parties to join the project. 

Moreover, the XShooter data is open to the public and we also plan to make the higher level data products open to the community at large, as these high-quality data have long-term utility for research projects that may not as yet have been foreseen.

\onecolumn
\begin{footnotesize}
\begin{landscape}
\setlength\LTcapwidth{\linewidth}
\begin{longtable}{l@{\hspace{2mm}}l@{\hspace{2mm}}l@{\hspace{2mm}}c@{\hspace{2mm}}c@{\hspace{2mm}}c@{\hspace{2mm}}l@{\hspace{2mm}}c@{\hspace{2mm}}c@{\hspace{2mm}}c@{\hspace{2mm}}l@{\hspace{2mm}}c@{\hspace{2mm}}c@{\hspace{2mm}}l@{\hspace{2mm}}c@{\hspace{2mm}}l@{\hspace{2mm}}l@{\hspace{2mm}}l}
\caption{Photometric, spectroscopic and multiplicity of ULLYSES/XShootU targets, ordered by RA (Magellan/MIKE targets are indicated in the final column). The infrared photometry is drawn from 2MASS ($JHK_s$) or VISTA VMC ($JK_s$). 
$H$-band photometry is missing if 2MASS and VISTA are discrepant (presumably due to photometric variability or sensitivity/crowding issues).} \label{table:targets}\\
\hline\hline
Target, Alias & Sp Type & Ref &	$U$ &    $B$ &	   $V$ & Ref & $J$    & $H$    & $K_s$ & Ref & $(B-V)_0$ & $E(B-V)$ & Ref & $M_{V}$  & Binary? & Ref & MIKE?\\ 
                     &              &        &(mag)  & (mag)  & (mag) &       & (mag) & (mag)  & (mag) &        & (mag)            & (mag)       &       & (mag) & & &\\  
\hline
\endfirsthead
\multicolumn{4}{c}%
{\tablename\ \thetable\ -- \textit{Continued from previous page}} \\
\hline
Target, Alias & Sp Type & Ref &	$U$ &    $B$ &	   $V$ & Ref & $J$    & $H$    & $K_s$ & Ref & $(B-V)_0$ & $E(B-V)$ & Ref & $M_{V}$  & Binary? & Ref & MIKE?\\ 
                     &              &        &(mag)  & (mag)  & (mag) &       & (mag) & (mag) & (mag)  &        & (mag)            & (mag)       &       & (mag) & & &\\  
\hline
\endhead
\hline \multicolumn{4}{r}{\textit{Continued on next page}} \\
\endfoot
\hline
\endlastfoot
\multicolumn{18}{c}{\bf SMC targets} \\ 
2dFS 163                                       & O8 Ib(f)               & 1   & 13.84 & 14.95 & 15.11 &  2 & 15.28 & 15.31 & 15.14  & 3,4 & --0.27      & 0.11 & 5 & --4.21 & $\cdots$ & $\cdots$& $\cdots$ \\  
AzV 6, 2dFS 5002                          & O9 III                  & 6       & 12.45 & 13.36 & 13.31 &      7 & 12.56 & 11.78 & 11.73 & 3,8 & --0.26     & 0.31 & 5 & --6.63 & EB & 9 & $\cdots$\\  
AzV 14, Sk~9                                  & O5 V                  & 10  & 12.38 & 13.38 & 13.55 &     7 & 14.25 & 14.30 & 14.31 & 3,8 & --0.28     & 0.11 & 5 & --5.77 & $\cdots$ &$\cdots$& $\cdots$\\   
AzV 15, Sk~10                                & O6.5 II(f)            &  11 & 11.90 & 12.93 & 13.12 &     7  & 13.53  &13.57 & 13.60 & 3,8 & $\cdots$   & 0.10 & 12 & --6.17 & $\cdots$ &$\cdots$ & $\cdots$\\  
AzV 16, Sk~11, R~4                       & sgB0$[ \,$e$] \,$ & 13  & 12.29 & 13.10 & 12.97 &     7 & 12.35 & 11.73 & 11.35 & 3,8 &  $\cdots$ & 0.07 & 14 & --6.23 & SB2 & 14& $\cdots$\\  
AzV 18, Sk~13                               & B2 Ia                   & 6       & 11.70 & 12.49 & 12.46 &   15 & 12.47 &12.36 & 12.32 & 3,8 & --0.17     & 0.20 & 16        & --7.16  & $\cdots$ &$\cdots$ & $\checkmark$ \\
\hline
\end{longtable}
\begin{flushleft}
References: 
{\bf 1} \citet{2004MNRAS.353..601E}, 
{\bf 2} \citet{2002AJ....123..855Z},
{\bf 3} \citet{2011A&A...527A.116C},
{\bf 4} \citet{2012yCat.2281....0C},
{\bf 5} \citet{2006A&A...457..637M},
{\bf 6} \citet{1997A&A...317..871L}, 
{\bf 7} \citet{2002ApJS..141...81M},
{\bf 8} \citet{2003yCat.2246....0C},
{\bf 9} \citet{2017MNRAS.472.2241Z},
{\bf 10} \citet{2004ApJ...608.1001M},
{\bf 11} \citet{2000PASP..112.1243W},
{\bf 12} \citet{2021A&A...647A.134B},
{\bf 13} \citet{1998A&A...340..117L},
{\bf 14} \citet{1996A&A...309..505Z},
{\bf 15} \citet{1977A&AS...30..261A},
{\bf 16} \citet{1970A&A.....4..234F},
{\bf 17} \citet{1980A&AS...42....1A},
{\bf 18} \citet{2003MNRAS.338..360F},
{\bf 19} \citet{2015A&A...581A..21H},
{\bf 20} \citet{2013ApJ...763..101L},
{\bf 21} \citet{2016ApJ...817..113L},
{\bf 22} \citet{2002ApJS..141..443W},
{\bf 23} \citet{1975A&AS...22..285A},
{\bf 24} \citet{2004ApJ...610.1021E},
{\bf 25} \citet{2006A&A...457..949M},
{\bf 26} \citet{1982A&AS...50..291A},
{\bf 27} \citet{1995ApJ...438..188M},
{\bf 28} \citet{2013A&A...555A...1B},
{\bf 29} \citet{2006A&A...456..623E},
{\bf 30} \citet{1987AJ.....93.1070G},
{\bf 31} \citet{2005ApJ...627..477M},
{\bf 32} \citet{1995PASP..107..104W},
{\bf 33} \citet{1989AJ.....98.1305M},
{\bf 34} \citet{2004ApJ...608.1028W},
{\bf 35} \citet{2009ApJ...692..618M},
{\bf 36} \citet{2019A&A...626A..50D},
{\bf 37} \citet{1991A&A...243..401H},
{\bf 38} \citet{2002IAUS..207..202N}, %
{\bf 39} \citet{2016A&A...591A..22S},
{\bf 40} \citet{2014AJ....148...62K},
{\bf 41} \citet{2012A&A...537A..29R},
{\bf 42} \citet{1977ApJ...215...53W},
{\bf 43} \citet{1983ApJ...265..716W},
{\bf 44} \citet{2013AcA....63..323P},
{\bf 45} \citet{1982PASP...94...31C},
{\bf 46} \citet{2010AJ....139.1283W},
{\bf 47} \citet{1997AJ....114.1951S},
{\bf 48} \citet{2018A&A...616A.103S},
{\bf 49} \citet{2022A&A...659A...9P},
{\bf 50} \citet{1993MNRAS.261..337R}
{\bf 51} \citet{2019A&A...625A.104R},
{\bf 52} \citet{1969AJ.....74..877S},
{\bf 53} \citet{1991PASP..103.1123F},
{\bf 54} \citet{1999MNRAS.306..279S},
{\bf 55} \citet{2017AJ....154..102U},
{\bf 56} \citet{1972A&AS....6..249A},
{\bf 57} \citet{1988ApJ...335..703F},
{\bf 58} \citet{1978A&AS...31..243R},
{\bf 59} \citet{1996MNRAS.281..163S},
{\bf 60} \citet{2014A&A...565A..27H},
{\bf 61} \citet{1979A&AS...38..239I},
{\bf 62} \citet{1992AJ....103.1205P},
{\bf 63} \citet{2002AJ....123.2754W},
{\bf 64} \citet{2017ApJ...837..122M},
{\bf 65} \citet{2004AJ....128.1606Z},
{\bf 66} \citet{2011AcA....61..103G},
{\bf 67} \citet{1999RMxAC...8..149M},
{\bf 68} \citet{1990ApJ...348..471S},
{\bf 69} \citet{2002A&A...392..653C},
{\bf 70} \citet{2011MNRAS.416.1311C},
{\bf 71} \citet{2008MNRAS.389..806S},
{\bf 72} \citet{2000AJ....119.2214M},
{\bf 73} \citet{2002ApJ...579..774C},
{\bf 74} \citet{1995msct.book.....B},
{\bf 75} \citet{1986AJ.....92...48C},
{\bf 76} \citet{2009PASP..121..634B},
{\bf 77} \citet{1997AJ....114..326A},
{\bf 78} \citet{1987PASP...99..240G},
{\bf 79} \citet{1982A&AS...50....7I},
{\bf 80} \citet{1975A&AS...19..259I},
{\bf 81} \citet{1983A&AS...51..505F},
{\bf 82 }\citet{1997A&A...320..500C},
{\bf 83} Crowther (2021, priv.\ com.),
{\bf 84} \citet{2003MNRAS.338..141O},
{\bf 85} \citet{1994A&A...288L...9H},
{\bf 86} \citet{1998AJ....116.1263O}, 
{\bf 87} \citet{1994AJ....108.1256G},
{\bf 88} \citet{2018A&A...615A..40R},
{\bf 89} \citet{2018A&A...609A...7R},
{\bf 90} \citet{2015A&A...584A...5E},
{\bf 91} \citet{2002A&A...385..517N},
{\bf 92} \citet{1978ApJ...225..548H},
{\bf 93} \citet{2013AcA....63..159U},
{\bf 94} \citet{2020A&A...634A.118M},
{\bf 95} \citet{2011A&A...530A.108E},
{\bf 96} \citet{2014A&A...564A..40W},
{\bf 97} \citet{2013A&A...550A.107S},
{\bf 98} \citet{2022A&A...665A.148S},
{\bf 99} \citet{1993AJ....106..560P},
{\bf 100} \citet{1999A&A...341...98S},
{\bf 101} \citet{2019A&A...627A.151S},
{\bf 102} \citet{2015A&A...574A..13E},
{\bf 103} \citet{1998A&AS..130..527T},
{\bf 104} \citet{2009AJ....138..510F},
{\bf 105} \citet{2018MNRAS.474.1412G},
{\bf 106} \citet{2004NewAR..48..727N},
{\bf 107} \citet{2007ApJ...659.1198E},
{\bf 108} \citet{2022MNRAS.516.4164L},
{\bf 109} \citet{2007AJ....133.2393M},
\end{flushleft}
\end{landscape}
\end{footnotesize}
\twocolumn

\begin{landscape}
\begin{table}
\caption{Baseline LMC metal abundances (X/H by number) with respect to $Z_{\odot}$ from \citet[MBS22]{2022A&A...661A.140M}. 
Values shown in parentheses are excluded from the average, due to issues with enhancements owing to mixing (e.g., N), depletion on dust grains (Fe) or other concerns. $\log (P/H)+12 = 5.1$ or 0.5 $Z_{\odot}$ is adopted for P following ISM studies of \citet{2015ApJ...811...78T}.}
\label{LMCabundances}
\begin{center}
\begin{footnotesize}
\begin{tabular}{
l@{\hspace{1mm}} 
c@{\hspace{2mm}} 
c@{\hspace{1mm}}c@{\hspace{1mm}}c@{\hspace{1mm}}c@{\hspace{2mm}} 
c@{\hspace{1mm}}c@{\hspace{1mm}}c@{\hspace{2mm}}  
c@{\hspace{1mm}}c@{\hspace{2mm}} 
c@{\hspace{2mm}} 
c@{\hspace{2mm}} 
c@{\hspace{2mm}}c} 
\hline
Element & SNR& \multicolumn{4}{c}{{---}{---}{---} H\,{\sc ii} regions {---}{---}{---}} & \multicolumn{3}{c}{{---}{---} B stars {---}{---}} & \multicolumn{2}{c}{--- F supergiants --- } & <LMC>      & Solar   & \multicolumn{2}{c}{<LMC>}\\
              & D19 & G99 & P03/SCT17 & T03 & L08   & K00/K05 & T07/H07    & D18  & HAS95 & A01 & $X/H$ & MBS22 & $\log X/H+12$ & $Z_{\odot}$ \\
\hline
C & 1.2$\times 10^{-4}$ & 7.9$\times 10^{-5}$ & 1.0$\times 10^{-4}$ & $\cdots$ & $\cdots$ & 9.5$\times 10^{-5}$ & $\cdots$ & $\cdots$ & 1.2$\times 10^{-4}$ & 8.9$\times 10^{-5}$ & 1.0$\times 10^{-4}$ & 3.6$\times 10^{-4}$ & 8.01 & 0.28 \\
N & (1.6$\times 10^{-5}$) & 7.9$\times 10^{-6}$ & 1.6$\times 10^{-5}$ & $\cdots$ & $\cdots$ & 1.0$\times 10^{-5}$   & 7.9$\times 10^{-6}$ & $\cdots$ & $\cdots$ & $\cdots$ & 1.1$\times 10^{-5}$ & 9.5$\times 10^{-5}$ & 7.03 & 0.12 \\
O & 2.2$\times 10^{-4}$ & 2.5$\times 10^{-5}$ & 3.2$\times 10^{-4}$ & 2.2$\times 10^{-4}$ & $\cdots$ & 2.5$\times 10^{-4}$ & 2.3$\times 10^{-4}$ & $\cdots$ & (2.8$\times 10^{-4}$) & (3.2$\times 10^{-4}$) & 2.5$\times 10^{-4}$ & 5.9$\times 10^{-4}$ & 8.40 & 0.42 \\
Ne & 4.0$\times 10^{-5}$ &  4.0$\times 10^{-5}$ & 6.8$\times 10^{-5}$ & 4.6$\times 10^{-5}$ & 5.8$\times 10^{-5}$ & $\cdots$ & $\cdots$ & $\cdots$ & $\cdots$ & $\cdots$ & 5.0$\times 10^{-5}$ & 1.4$\times 10^{-4}$ & 7.70 & 0.35 \\
Na & $\cdots$ & $\cdots$ & $\cdots$ & $\cdots$ & $\cdots$ & $\cdots$ & $\cdots$ & $\cdots$ & 1.5$\times 10^{-6}$ & 7.9$\times 10^{-7}$ & 1.2$\times 10^{-6}$ & 1.9$\times 10^{-6}$ & 6.1\phantom{0} & 0.6\phantom{0} \\
Mg & 1.5$\times 10^{-5}$ & $\cdots$ & $\cdots$ &  $\cdots$ & $\cdots$ & 1.4$\times 10^{-5}$ & 1.2$\times 10^{-5}$ & 1.0$\times 10^{-5}$ & 1.6$\times 10^{-5}$ & $\cdots$ & 1.3$\times 10^{-5}$ & 3.5$\times 10^{-5}$ & 7.13 & 0.46 \\
Al & $\cdots$ & $\cdots$ & $\cdots$ &  $\cdots$ & $\cdots$ & 7.2$\times 10^{-7}$ & $\cdots$ & $\cdots$ & 2.2$\times 10^{-6}$ & $\cdots$ & 1.5$\times 10^{-6}$ & 2.7$\times 10^{-6}$ & 6.2\phantom{0} & 0.5\phantom{0} \\
Si & 1.3$\times 10^{-5}$ &   5.0$\times 10^{-6}$ & $\cdots$ & $\cdots$ & $\cdots$ & 1.2$\times 10^{-5}$ & 1.6$\times 10^{-5}$ & $\cdots$ & (3.7$\times 10^{-5}$)  & $\cdots$ & 1.1$\times 10^{-5}$ & 3.9$\times 10^{-5}$ & 7.06 & 0.32 \\
S & 9.5$\times 10^{-6}$ &  5.0$\times 10^{-6}$ & 9.8$\times 10^{-6}$ & 5.9$\times 10^{-6}$ & 5.9$\times 10^{-6}$ & $\cdots$ & $\cdots$ & $\cdots$ & $9.9\times 10^{-6}$ & $\cdots$ & 7.7$\times 10^{-6}$ & 1.4$\times 10^{-5}$ & 6.88 & 0.53 \\
Cl & 9.1$\times 10^{-8}$ &  $\cdots$  & 6.6$\times 10^{-8}$ & $\cdots$ & $\cdots$ & $\cdots$ & $\cdots$ & $\cdots$ & $\cdots$ & $\cdots$ & 7.9$\times 10^{-8}$ & 1.8$\times 10^{-7}$ & 4.9\phantom{0} & 0.4\phantom{0} \\ 
Ar & 6.2$\times 10^{-7}$ &   1.6$\times 10^{-6}$ & 1.8$\times 10^{-6}$ & 1.4$\times 10^{-6}$ & 2.1$\times 10^{-6}$ & $\cdots$ & $\cdots$ & $\cdots$ & $\cdots$ & $\cdots$ & 1.5$\times 10^{-6}$ & 3.2$\times 10^{-6}$ & 6.2\phantom{0} & 0.5\phantom{0} \\
Ca & 1.0$\times 10^{-6}$ & $\cdots$ & $\cdots$ & $\cdots$ & $\cdots$ & $\cdots$ & $\cdots$ & $\cdots$ & 1.6$\times 10^{-6}$ & $\cdots$ & 1.3$\times 10^{-6}$ & 2.3$\times 10^{-6}$ & 6.1\phantom{0} & 0.6\phantom{0} \\
Cr & 2.2$\times 10^{-7}$ & $\cdots$ & $\cdots$ & $\cdots$ & $\cdots$ & $\cdots$ & $\cdots$ & $\cdots$ & 2.6$\times 10^{-7}$ & $\cdots$ & 2.4$\times 10^{-7}$ & 5.5$\times 10^{-7}$ & 5.4\phantom{0} & 0.4\phantom{0} \\
Fe & 2.1$\times 10^{-5}$ &  $\cdots$ & (2.5$\times 10^{-6}$)  & $\cdots$ & (1.7$\times 10^{-6}$) & 1.2$\times 10^{-5}$ & 1.7$\times 10^{-5}$ & $\cdots$ & 1.7$\times 10^{-5}$ & $\cdots$ & 1.7$\times 10^{-5}$ & 3.2$\times 10^{-5}$ & 7.23 & 0.54 \\
Ni & 7.9$\times 10^{-7}$  & $\cdots$ & $\cdots$ & $\cdots$ & $\cdots$ & $\cdots$ & $\cdots$ & $\cdots$ & 9.2$\times 10^{-7}$ & $\cdots$ & 8.6$\times 10^{-7}$ & 1.7$\times 10^{-6}$ & 5.9\phantom{0} & 0.5\phantom{0} \\
\hline
\end{tabular}
\end{footnotesize}
\end{center} 
HAS95: \citet{1995A&A...293..347H};
G99: \citet{1999IAUS..190..266G}; 
K00: \citet{2000A&A...353..655K} (He/Al/Fe only); 
K05: \citet{2005ApJ...633..899K}; (excl He/Al/Fe) 
A01: \citet{2001A&A...367..605A}; 
P03: \citet{2003ApJ...584..735P} (excl. C/O);
T03: \citet{2003MNRAS.338..687T}; 
T07: \citet{2007A&A...471..625T}; 
H07: \citet{2007A&A...466..277H}; 
L08: \citet{2008ApJ...680..398L}; 
SCT17: \citet{2017MNRAS.467.3759T} (C/O recomb lines for 30~Dor); 
D18: \citet{2018A&A...615A.101D}; 
D19: \citet{2019AJ....157...50D} 
\end{table}
\end{landscape}


\begin{landscape}
\begin{table}
\caption{Baseline SMC metal abundances (X/H by number) adopted with respect to $Z_{\odot}$ from \citet[MBS22]{2022A&A...661A.140M}. 
Values shown in parentheses are excluded from the average, due to issues with enhancements owing to mixing (e.g., N), depletion on dust grains (Fe) or other concerns. $\log(P/H)+12=4.7$ or 0.2 $Z_{\odot}$ is adopted for P from ISM studies of \citet{2015ApJ...811...78T}.}
\label{SMCabundances}
\begin{center}
\begin{tabular}{
l@{\hspace{1mm}}
c@{\hspace{2mm}} 
c@{\hspace{1mm}}c@{\hspace{1mm}}c@{\hspace{1mm}}c@{\hspace{2mm}} 
c@{\hspace{2mm}} 
c@{\hspace{1mm}}c@{\hspace{2mm}} 
c@{\hspace{1mm}}c@{\hspace{2mm}} 
c@{\hspace{2mm}} 
c@{\hspace{2mm}} 
c@{\hspace{2mm}}c} 
\hline
Element & SNR & \multicolumn{4}{c}{{---}{---}{---} H\,{\sc ii} regions {---}{---}{---}} & A stars & \multicolumn{2}{c}{--- B stars ---} & \multicolumn{2}{c}{ K supergiants} & <SMC>      & Solar   & \multicolumn{2}{c}{<SMC>} \\
              & D19 & KD98  & G99 & R02      & L08                      & V99     &  K00 & T07/H09      & H97/HBS97 & H99 & $X/H$ & MBS22 & $\log X/H+12$ & $Z_{\odot}$ \\
\hline
C   & 3.2$\times 10^{-5}$ & 1.6$\times 10^{-5}$ & 2.5$\times 10^{-5}$ & $\cdots$                  & $\cdots$                   & $\cdots$                    & 2.5$\times 10^{-5}$    & 2.0$\times10^{-5}$    & 3.6$\times 10^{-5}$    & 2.8$\times 10^{-5}$ & 2.6$\times 10^{-5}$ & 3.6$\times 10^{-4}$ & 7.42 & 0.07 \\
N   & 6.6$\times 10^{-6}$ & 3.6$\times 10^{-6}$ & 3.2$\times 10^{-6}$ & 6.5$\times 10^{-6}$ & $\cdots$                   & (2.1$\times 10^{-5}$) & 3.2$\times 10^{-6}$  & (1.7$\times 10^{-5}$) & (3.2$\times 10^{-5}$) & $\cdots$                 & 4.6$\times 10^{-6}$ & 9.5$\times 10^{-5}$ & 6.66 & 0.05 \\
O  & 1.0$\times 10^{-4}$  & 1.0$\times 10^{-4}$ & 1.0$\times 10^{-4}$ & 1.4$\times 10^{-4}$ & $\cdots$                   & 1.4$\times 10^{-4}$  & 1.4$\times 10^{-4}$   & 9.8$\times 10^{-5}$   & 1.2$\times 10^{-4}$    & $\cdots$                 & 1.1$\times 10^{-4}$ & 5.9$\times 10^{-4}$ & 8.05 & 0.19 \\
Ne & 1.1$\times 10^{-5}$ & 1.6$\times 10^{-5}$ &  1.6$\times 10^{-5}$ & 2.1$\times 10^{-5}$ & 2.2$\times 10^{-5}$ & $\cdots$                    &  $\cdots$                  & $\cdots$                     & 2.9$\times 10^{-5}$    & $\cdots$                 & 1.7$\times 10^{-5}$ & 1.4$\times 10^{-4}$ & 7.23 & 0.12 \\
Na & $\cdots$                  &  $\cdots$                 &                                  & $\cdots$.                  & $\cdots$                  & $\cdots$                    & $\cdots$                  & $\cdots$                      & 4.0$\times 10^{-7}$    &  2.6$\times 10^{-7}$ & 3.1$\times 10^{-7}$& 1.9$\times 10^{-6}$ & 5.49 & 0.16\\
Mg & 5.2$\times 10^{-6}$ & $\cdots$                 & $\cdots$                    &  $\cdots$                 & $\cdots$.                 & 6.7$\times 10^{-6}$   & 6.0$\times 10^{-6}$ & 5.2$\times 10^{-6}$    & 1.1$\times 10^{-5}$     & 4.5$\times 10^{-6}$ & 6.1$\times 10^{-6}$ & 3.5$\times 10^{-5}$ & 6.78 & 0.17 \\
Al & $\cdots$                   & $\cdots$                  & $\cdots$                    &  $\cdots$                 & $\cdots$                  & $\cdots$                    & 3.8$\times 10^{-7}$ & $\cdots$                     & 4.6$\times 10^{-7}$    & 3.6$\times 10^{-7}$ & 3.9$\times 10^{-7}$ & 2.7$\times 10^{-6}$ & 5.59 & 0.15 \\
Si & 5.9$\times 10^{-6}$ & $\cdots$                  & 2.0$\times 10^{-6}$      &  $\cdots$                 & $\cdots$                  & 9.3$\times 10^{-6}$   & 6.9$\times 10^{-6}$ & 5.9$\times 10^{-6}$    & 5.2$\times 10^{-6}$    & 5.6$\times 10^{-6}$ & 5.3$\times 10^{-6}$ & 3.9$\times 10^{-5}$ & 6.72 & 0.14 \\
S & $\cdots$                   & $\cdots$                   & 2.0$\times 10^{-6}$     & 2.5$\times 10^{-6}$ & 2.3$\times 10^{-6}$ & $\cdots$                    & $\cdots$                  & $\cdots$                    & $\cdots$.                     & $\cdots$                  & 2.3$\times 10^{-6}$ & 1.4$\times 10^{-5}$ & 6.36 & 0.16 \\
Cl & 2.3$\times 10^{-8}$ & $\cdots$.                  & $\cdots$                   & $\cdots$                  & $\cdots$                  & $\cdots$                    & $\cdots$                  & $\cdots$                    & $\cdots$                       & $\cdots$                  &2.3$\times 10^{-8}$ &  1.8$\times 10^{-7}$ & 4.37 & 0.13 \\
Ar & 5.1$\times 10^{-7}$ & $\cdots$                   &  7.9$\times 10^{-7}$ & 6.6$\times 10^{-7}$ & 9.3$\times 10^{-7}$ & $\cdots$                   & $\cdots$                  & $\cdots$                     & $\cdots$                      & $\cdots$                 & 7.2$\times 10^{-7}$ & 3.2$\times 10^{-6}$ & 5.86 & 0.23 \\
Ca & 5.9$\times 10^{-7}$ & $\cdots$                  & $\cdots$                   & $\cdots$                 & $\cdots$                    & 2.1$\times 10^{-7}$  & $\cdots$                   & $\cdots$                   & 3.9$\times 10^{-7}$    & 3.3$\times 10^{-7}$ & 4.1$\times 10^{-7}$ & 2.3$\times 10^{-6}$ & 5.61 & 0.17 \\ 
Cr & $\cdots$                   & $\cdots$                  & $\cdots$                   & $\cdots$                 & $\cdots$                     & $\cdots$                  &  $\cdots$                 & $\cdots$                    & 4.6$\times 10^{-8}$    & 3.7$\times 10^{-8}$ & 4.0$\times 10^{-8}$ & 5.5$\times 10^{-7}$ & 4.61 & 0.07 \\ 
Fe & 5.9$\times 10^{-6}$ & $\cdots$                  & $\cdots$                   &  $\cdots$                &  $\cdots$                    & 5.1$\times 10^{-6}$ & 6.6$\times 10^{-6}$ & 8.5$\times 10^{-6}$ & 6.4$\times 10^{-6}$   & 5.1$\times 10^{-6}$ & 6.3$\times 10^{-6}$ & 3.2$\times 10^{-5}$ & 6.80 & 0.20 \\
Ni & 3.5$\times 10^{-7}$  & $\cdots$                  & $\cdots$                  & $\cdots$                  & $\cdots$                    & $\cdots$                   & $\cdots$                  & $\cdots$                   & 3.4$\times 10^{-7}$    & 1.2$\times 10^{-7}$ & 2.4$\times 10^{-7}$ & 1.7$\times 10^{-6}$ & 5.37 & 0.14 \\
\hline
\end{tabular}
\end{center}
H97: \citet{1997A&A...324..435H}
HBS97; \citet{1997A&A...323..461H}
KD98: \citet{1998RMxAC...7..202K}
G99: \citet{1999IAUS..190..266G} 
V99: \citet{1999ApJ...518..405V} 
H99: \citet{1999A&A...345..430H}
K00: \citet{2000A&A...353..655K}
T07: \citet{2007A&A...471..625T} 
L08: \citet{2008ApJ...680..398L} 
H09: \citet{2009A&A...496..841H}
D19: \citet{2019AJ....157...50D} 
\end{table}
\end{landscape}

\begin{acknowledgements}
We thank the ESO support staff for the help in the preparation of the observations, in the scheduling, and for carrying out the observations at Paranal. We also thank the ESO support staff, in particular Carlo Felice Manara.
We also thank the ULLYSES teams at STScI, in particular, Julia Roman-Duval and STScI Director Ken Sembach, for executing and enabling the ULLYSES Director's Discretionary program.
Based on observations obtained with the NASA/ESA Hubble Space Telescope, retrieved from the Mikulski Archive for Space Telescopes (MAST) at the Space Telescope Science Institute (STScI). STScI is operated by the Association of Universities for Research in Astronomy, Inc. under NASA contract NAS 5-26555.

JSV and ERH gratefully acknowledge support from STFC via grant ST/V000233/1.

LPM thanks CNPQ for financial support through grant 307115/2021-6

AW acknowledges the support of UNAM via grant agreement PAPIIT no. IN106922

ADU acknowledges support from NASA under award number 80GSFC21M0002.

DMB gratefully acknowledges funding from the Research Foundation Flanders (FWO) by means of a senior postdoctoral fellowship with grant agreement number 1286521N.

AACS acknowledges support by the Deutsche Forschungsgemeinschaft (DFG - German Research Foundation) 
in the form of an Emmy Noether Research Group -- Project-ID 445674056 (SA4064/1-1, PI Sander).

BK gratefully acknowledges support from the Grant Agency of the Czech Republic (GA\v CR 22-34467S). The Astronomical Institute in Ond\v rejov is supported by the
project RVO:67985815.

CJE gratefully acknowledges support for this work provided by NASA through grant number {\it HST-AR-15794.001-A} from the Space Telescope Science Institute, which is operated by AURA, Inc., under NASA contract NAS 5-26555.

NSL wishes to thank the National Sciences and Engineering Council of Canada (NSERC) for financial support.

AuD acknowledges support by NASA through Chandra Award number TM1-22001B and GO2-23003X issued by the Chandra X-ray Observatory 27 Center, which is operated by the Smithsonian Astrophysical Observatory for and on behalf of NASA under contract NAS8-03060. In  addition, AuD acknowledges NASA ATP grant number 80NSSC22K0628.

GM acknowledges funding support from the EuŁropean Research Council (ERC) under the European Union’s HoriŁzon 2020 research and innovation programme (Grant agreement No.Ł 772086).

RI gratefully acknowledges support by the National Science Foundation under Grant No. AST-2009412.

JMA acknowledges support from the Spanish Government Ministerio de Ciencia e Innovaci\'on through grant PGC2018-095\,049-B-C22. 

CJKL gratefully acknowledges support from the International Max Planck Research School for Astronomy and Cosmic Physics at the University of Heidelberg in the form of an IMPRS PhD fellowship.

NDK acknowledges support from the National Solar Observatory, which is managed by the Association of Universities for Research in Astronomy, Inc. and funded by the National Science Foundation.

MG and FN gratefully acknowledge funding by grants PID2019-105552RB-C41 and
MDM-2017-0737 Unidad de Excelencia ‘‘Mar\'{\i}a de Maeztu’’-Centro de Astrobiolog\'{\i}a
(INTA-CSIC) by the Spanish Ministry of Science and Innovation/ State Agency
of Research MCIN/AEI/10.13039/501100011033 and by “ERDF A way of
making Europe”.

\end{acknowledgements}

\appendix

\section{The XShootU collaboration}
\label{sect:organisation}

\subsection{Organisation (WG\,1)}

XShootU is a project with a wide-scale community approach. The collaboration is organised in 14 working groups (WGs) that anyone can join: 
\begin{enumerate}
    \item {Organisation}
    \item {Data Reduction and Calibration}
    \item {Stellar Atmospheres and Benchmarking}
    \item {Wind Structure}
    \item {Stellar Evolution Modelling}
    \item {Stellar Libraries}
    \item {Population Synthesis}
    \item {Interstellar Medium}
    \item {Massive Star Feedback}
    \item {Stripped Stars}
    \item {Auxiliary/New Data}
    \item {Pulsations}
    \item {Magnetic Fields}
    \item {Unusual Objects}
\end{enumerate}

\subsection{Data Reduction and Calibration (WG\,2)}\label{WG2}

The detailed description of the UVB and VIS higher level data products (HLDPs) will be associated with Data Release 1 (DR1; Sana et al.\ 2023, XShootU {\sc II}). The data reduction of the NIR spectra will become part of DR2.

\subsection{Stellar Atmospheres and Benchmarking (WG\,3)}\label{WG3}

WG\,3 was set up to foster discussions and cooperation between researchers performing the spectroscopic analysis of the XShootU datasets. Part of this work is described in the main part of the paper.
Several sub-groups were organised by WG\,3 members to address specific science questions and topics of interest.
This ranges from the determination of the full photospheric and wind properties of the sample at large, to very specific issues such as studying the effects of rotational mixing on stars with different masses/ages through abundance studies. 
The three main codes that are employed for the  spectroscopic analysis are the non-LTE expanding wind codes CMFGEN, FASTWIND, and PoWR \citep{1998ApJ...496..407H, 2020A&A...642A.172P, 2002A&A...387..244G}, and the plane parallel non-LTE code TLUSTY for stars without strong winds.

\subsection{Wind Structure (WG\,4)}

Winds of massive, hot stars are time-dependent and highly structured on small and large spatial scales \citep[see overviews in][]{Puls:2008, Hamann:2008}. It has been shown that properties of structured winds have an enormous impact on empirical estimates of mass-loss rates, and thus on our understanding of stellar evolution. However, the question to which extent $\dot{M}$ is influenced by wind structures remains. 
WG\,4 studies the clumping dependence on spectral and luminosity classes at different $Z$. Spectral modelling is performed using state-of-the-art model atmosphere codes (see WG\,3), which can treat clumping properties with different levels of sophistication.
Wind clumping is usually treated in the ``microclumping'' (i.e., optically thin clumps at all frequencies) or ``macroclumping'' (i.e., arbitrary optical thickness of clumps) approximations, still assuming a void inter-clump medium and a smooth velocity field (e.g., CMFGEN, PoWR, and METUJE -- \citealt{Krticka:Kubat:2017,Krticka:Kubat:2018}).

\subsection{Stellar Evolution Modelling (WG\,5)}

The objectives of the stellar evolution group fall into two categories. The first one is to provide up to date sets of stellar evolution models covering not only the parameter space of initial conditions but also that of uncertain physical prescriptions such as overshooting. Following the work of \citet{Schneider2014} we will also provide open source tools\footnote{An early version of these tools can be found at \url{https://github.com/orlox/StarStats.jl}} to perform Bayesian inference of observed systems and infer their initial properties given a set of simulations. 
Even though the XShootU sample was designed to exclude known binary stars, it is possible that such a sample contains large numbers of binary products. Therefore extended grids of binary evolution will also be constructed and incorporated into the Bayesian inference framework under construction. Our simulations will be coded using the MESA software instrument \citep{Paxton2011,Paxton2013,Paxton2015,Paxton2018,Paxton2019}, but our Bayesian tools are designed to easily incorporate simulations from other stellar evolution simulations, allowing us to establish systematic biases on our theoretical models. The second objective of the working group will be the development of tailored evolutionary models for specific objects of interest found within the XShootU sample.

\subsection{Stellar Libraries (WG\,6)}

\begin{figure}
    \centering
    \includegraphics[width=0.49\textwidth]{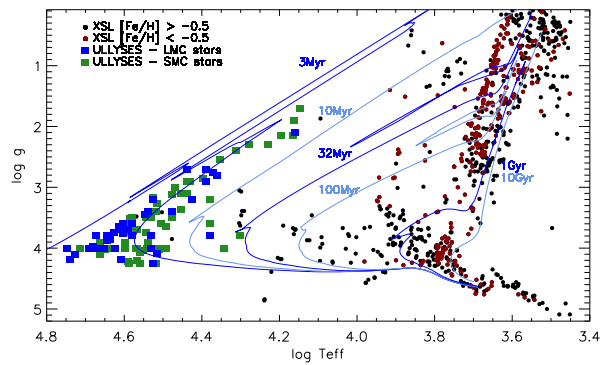}
    \caption{Comparison of the ULLYSES (blue: LMC; green: SMC) and XSL libraries.
XSL stars (filled circles) are separated by metallicity: [Fe/H] $>-0.5$ (black) and 
[Fe/H] $<-0.5$ (dark red). Temperatures, gravities, and metallicities of the ULLYSES stars in this figure were obtained from the literature. Solid lines are Geneva isochrones for $Z_\odot$ and various ages \citep{ekstrom+12}.}
        \label{fig_lmartins}
\end{figure}

Stellar spectral libraries are one of the main ingredients of stellar population synthesis models, which in turn are powerful tools in the study of fundamental properties of unresolved stellar
systems.
One of the major deficiencies of the empirical libraries available today is the coverage of hot and young stars at low $Z$ (Hill et al. 2022). 
A variety of empirical stellar spectral libraries, collected with different primary goals, are publicly available (see Table 1 in Verro et al. 2021 for a recent summary).
In terms of spectral resolving power, only the X-shooter Spectral Library (XSL) and ELODIE are comparable to the XShootU data-set.
ULLYSES $+$ XShootU is by far the most complete, highest-S/N, and
highest resolution library of hot, massive stars with the broadest
spectral coverage. 
While numerous libraries for low-mass stars
are available, these libraries are incomplete at high masses. 
To illustrate this point, we compare the ULLYSES $+$ 
XShootU target
sample with the XSL library \citep{2021arXiv211010188V} which was
designed for stellar population synthesis. 
In Figure~\ref{fig_lmartins}, we show the
HR diagram coverage of the XSL library which shows the lack
of massive OB stars at any Z. ULLYSES $+$ XShootU perfectly
complement the missing parameter space of the XSL library. In
combination, the two libraries allow self-consistent population
synthesis models (see WG\,7) of systems hosting both young and old stars.

\subsection{Population Synthesis (WG\,7)}

Population synthesis models are called semi-empirical if the stellar evolution tracks are theoretical but the individual stellar spectra are observed, and fully theoretical when both are calculated. 
XShootU observations will enable improvements of semi-empirical models because they will provide the most complete spectral library to date.
In addition, the XShootU observations will guide new generations of atmosphere and evolution models, which will help improve fully theoretical population synthesis models.
The new empirical library of LMC $+$ SMC UV-to-NIR massive-star spectra built by the ULLYSES and XShootU projects will replace the Starburst99 LMC $+$ SMC library, greatly improving the realism of population synthesis predictions. 
Once the library is implemented in Starburst99, corresponding Cloudy \citep{2017RMxAA..53..385F} photoionization models will be computed in order to account for the contribution of the ionized gas and dust to the integrated light of young OB star populations. The models, which will be publicly available, will be used to interpret existing and future observations of star-forming and starburst galaxies with similar metallicities, in particular those from the COS Legacy Spectroscopic SurveY (CLASSY, \citealt{Berg22,2022arXiv220601224J}), which obtained the first
high-resolution ($R\sim15\,000$) FUV spectral database of 45 nearby ($0.002<z<0.182$) star-forming galaxies, the majority of which have LMC- or SMC-like metallicities.

\subsection{Interstellar Medium (WG\,8)}

The intervening ISM along the line of sight to each target star will impact the observed UV and optical spectra. Cool atomic and molecular gas produces absorption features superimposed on  stellar photospheric and wind lines. Furthermore, dust in the Milky Way and the host galaxy extinguishes and reddens the stellar spectra. These effects must be accounted for to reliably analyse stellar features, but also enable ISM science that is interesting in its own right.
Due to the nature of the UV-based selection process of the target sample and the fact that most OB stars in the MCs experience low extinctions ($E(B-V) < 0.20$ mag), strong ISM signatures are not expected in the optical and NIR X-shooter spectra. Nevertheless, clear signatures are expected in strong atomic lines such as the Na\,{\sc i}~$\lambda\lambda$5889.951,5895.924 and Ca\,{\sc ii}~$\lambda\lambda$3933.663,3968.468 doublets. For the targets with the strongest extinctions, detection are expected for some molecular lines, such as CH~$\lambda$4300.313 and CH+~$\lambda$4232.548, and strong diffuse interstellar bands (DIBs), such as those at 4428~\AA, 5780.5~\AA, 5797.1~\AA\ and 6614~\AA. These lines have been well studied for Galactic targets but less so for objects in the MCs \citep[though see, e.g.,][]{ 2002ApJ...576L.117E,2006ApJS..165..138W,2006A&A...447..991C,2007A&A...470..941C,2013A&A...550A.108V,2015MNRAS.454.4013B}. The coverage of the significantly less studied wavelength region between $3200 - 3800$~\AA\ could provide detections of some lines such as Ti\,{\sc i}~$\lambda$3729.8069 and Ti\,{\sc ii}~$\lambda$3383.759. In addition, the flux calibration of the X-shooter spectra will provide insights into the MC extinction laws \citep{Maizetal14a} and possibly detect the recently discovered very broad absorption band centered on 7700~\AA\ \citep{Maizetal21a}.

\subsection{Massive Star Formation and Feedback (WG\,9)}

The energy output from young OB stars creates expanding \ion{H}{ii} regions that are overpressurised with respect to their surroundings by photoionization, radiation pressure on dust grains, and shock heating by stellar winds \citep{Mathews1967, Weaver1977, SpitzerLyman1978}. 
Key stellar properties for the structure and dynamics of these regions are the flux in FUV and EUV photons and the stellar wind mass-loss rate and terminal velocity.
The density distribution, chemistry, and equilibrium temperature of the photoionised gas also play important roles \citep{Ferland2017}. 
Because the LMC and SMC have substantially lower metallicities than the Galaxy, their \ion{H}{ii} regions are ideal test cases for the impacts of stellar ionizing flux \citep{Voges2008, Oey1997} and winds \citep[e.g.,][]{Oey1996, Oey1998} in low-metallicity environments typical of high-redshift galaxies.
The XShootU dataset in the Magellanic Clouds represents a significant opportunity to advance the state of the art in our understanding of how stars shape their environment. 
WG9 has identified a sample of relatively isolated XShootU stars that reside inside \ion{H}{ii} regions identified in the \textsc{mcels} survey \citep{Pellegrini2012}. 
These single-star \ion{H}{ii} regions are valuable tools for calibrating stellar atmosphere models (like those produced by WG\,3) by comparing observed nebular emission lines to photoionization models adopting those model atmospheres as the ionizing sources \citep{Zastrow2013}.
Combining stellar properties inferred from the XShootU spectra with the observed structure and emission of the ionized gas in these regions will clarify the role stars play in shaping their environment.

\subsection{Stripped Stars (WG\,10)}

Stars stripped of their hydrogen-rich envelopes via mass transfer or common envelope ejection in binary systems are thought to be the fate of a third of all massive stars \citep{2012Sci...337..444S}. In fact, in a continuously star-forming population, a few percent of all massive stars should be accompanied by a stripped companion. Binary-stripped stars are the exposed helium cores of their progenitors. During the long-lived phase of core helium burning these stars are small ($R_\star \lesssim 1~R_{\odot}$), hot ($T_{\rm eff} \sim 50-100$~kK), compact ($\log g \sim 5$), helium-rich ($Y_{\rm surf} \sim 0.5-1$), and hydrogen-poor ($X_{\rm surf} \sim 0-0.5$), but span a range of luminosities ($1-10^6~L_\odot$)
and masses ($\sim 0.5-10~M_\odot$) \citep{2018A&A...615A..78G}. After core-helium exhaustion, they can expand to giant sizes, depending on the fraction of hydrogen they retain, and appear as helium-enriched blue, yellow, or red supergiants \citep{2017ApJ...840...10Y,2020A&A...637A...6L}.
ULLYSES offers an excellent sample to search for hot binary-stripped stars that are sufficiently massive to explode as SNe. Especially with the effort to exclude apparent binaries from the target sample, the likelihood of having a few hidden binary-stripped companions increases, as they are not expected to significantly contribute to the optical flux or to cause large radial velocities. 
While UV spectra will be crucial for this analysis, the optical spectra are needed for stellar characterisation of both the binary-stripped star and the accretor star. We expect that discovering binary-stripped star $+$ OB star systems in the ULLYSES sample will help constrain uncertainties, such as how common and efficient mass transfer is, and exactly how the two stars are affected by such mass transfer.

\subsection{Auxiliary/New Data (WG\,11)}

There are multiple ways in which auxiliary data can help to increase the value of the ULLYSES/XShootU library, including different wavelength regimes (e.g. X-rays) 
and observing strategies. 
For example, the intermediate spectral resolving power of the X-shooter data only partially resolves spectral lines, especially for stars with low projected rotational velocities (\vsini), and higher resolution spectra are needed for a detailed spectroscopic analysis.
WG\,11 will provide high-resolution optical spectroscopy of a selection of slow rotating  ULLYSES targets. Targets were selected on the basis of showing narrow, resolved C\,{\sc iii} 1176~\AA\,  lines in their UV spectra or having low \vsini\ reported in the literature.
In Dec 2021 - Dec 2022, we obtained high resolution ($R \sim 40\,000$) and high signal-to-noise ratio (S/N $\sim 200$ in the 4100~\AA\, region) optical spectra with the 6.5m Magellan Clay telescope (+MIKE spectrograph) for 48 ULLYSES targets. 
Basic reductions confirmed the high quality of the data, and final reductions are underway. The Magellan/MIKE data will be made available to the massive stars community.
WG11 also plans to organize archival searches aimed to identify data that will enhance the ULLYSES/XShootU library:
\begin{itemize}
    \item High spectral resolution optical spectra of ULLYSES targets, e.g., obtained with the UVES, FLAMES, FEROS, and HARPS instruments.
    \item UV spectra obtained with {\it IUE\/}, {\it FUSE\/}, and {\it  HST\/}.
\end{itemize}

\subsection{Pulsations (WG\,12)}

The study of spectroscopic variability in XShootU targets is naturally supported by the analysis of time-series photometry from the NASA Transiting Exoplanet Survey Satellite (TESS) mission \citep{2015JATIS...1a4003R}. The LMC is within the southern continuous viewing zone of the TESS mission, such that all XShootU stars in the LMC have continuous light curves spanning 1~yr. TESS light curves have been used to demonstrate that heat-driven pulsations and the signatures of rotational modulation are common in early-type main-sequence stars in the Galaxy \citep{2019MNRAS.487..304D, 2020A&A...639A..81B}. Whereas, the brightest massive stars in the MCs, i.e., OB supergiants, have light curves dominated by stochastic low-frequency variability (SLF; \citealt{2019NatAs...3..760B}). Interestingly, in low metallicity environments such as the MCs, the driving of heat-driven pulsations is less efficient \citep{2012MNRAS.422.3460S}. For massive stars with identified pulsation modes, forward asteroseismic modelling is able to provide constraints on their (core) masses, ages, interior rotation profiles, and mixing properties (e.g. \citealt{2003Sci...300.1926A, 2004A&A...415..251D, 2021arXiv211204064S}), but also probe the strength and geometry of an interior magnetic field \citep{2022MNRAS.512L..16L}. A key goal of WG12 is to extract reliable TESS light curves, provide complementary constraints on the variability amplitudes and timescales of each XShootU target, and ultimately perform forward asteroseismic modelling (see \citealt{2020FrASS...7...70B} for a review).

\subsection{Magnetic Fields (WG13)}

WG13 seeks to leverage the UV and optical spectral libraries provided by the ULLYSES and XShootU programmes to shed light on how magnetism affects massive star structure and evolution. In magnetic massive stars, which make up $\sim$7\% of the total OB star population \citep[e.g.,][]{Morel2015,Wade2016,Grunhut2017}, stable, nearly dipolar surface magnetic fields warp the stellar wind into a structurally complex magnetosphere, trapping the wind near the stellar surface, thus significantly altering the circumstellar environment. This {\it magnetic wind confinement} has been shown to reduce stellar mass-loss rate compared to non-magnetic stars of similar spectral type \citep{ud-Doula2008}.
It should be noted that atypical features in P-Cygni profiles are not necessarily the result of a magnetic fields, as  
variability in the absorption components of wind-sensitive UV resonance lines have also been observed in non-magnetic massive stars
\cite[e.g.,][]{Massa1995,Kaper1996,Kaper1999}. 
\citet{Erba2021b_uvadm}'s synthetic line profiles can be adapted for direct comparison with observations, and used in conjunction with the XShootU spectral libraries to identify new magnetic candidates.

\subsection{Unusual Objects (WG14)}

Finally, the data offer many opportunities to study key objects. 
For example, combining the radial velocities from XShootU with the exquisite proper motions from Gaia DR3, we can reconstruct the space motions of ULLYSES targets, which can be used to identify runaway and walkaway stars. 
Moreover, Gaia DR3 light-curves can be used in conjunction with radial velocities of known eclipsing binaries to estimate fundamental stellar parameters, providing unique tests of stellar evolution.

\section{ULLYSES target parameters}

In this Appendix we present a literature search for the stellar and wind parameters as they were known prior to the start of ULLYSES. Obviously, these data (see Table A1) are rather heterogeneous and they will quickly become outdated as new spectral analyses are underway.

\onecolumn
\begin{footnotesize}
\begin{landscape}
\setlength\LTcapwidth{\linewidth}
\begin{longtable}{l@{\hspace{2mm}}l@{\hspace{2mm}}l@{\hspace{2mm}}c@{\hspace{2mm}}c@{\hspace{2mm}}c@{\hspace{2mm}}l@{\hspace{2mm}}c@{\hspace{2mm}}c@{\hspace{2mm}}c@{\hspace{2mm}}l@{\hspace{2mm}}c@{\hspace{2mm}}c@{\hspace{2mm}}l@{\hspace{2mm}}c@{\hspace{2mm}}l@{\hspace{2mm}}c@{\hspace{2mm}}l@{\hspace{2mm}}c@{\hspace{2mm}}l@{\hspace{2mm}}c}
\caption{Pre-ULLYSES target parameters compiled by the ULLYSES team at STScI. References have been added in columns for all parameters.} \label{table:params}\\
\hline\hline
Target &	SpType & Ref &$T_\mathrm{eff}$ & Ref &	$\log~g$ &	Ref & $\log~ L/L_{\odot}$ & Ref &	$R/R_{\odot}$ &	Ref & $M/M_{\odot}$ & Ref &	\vsini & Ref &	$v_\mathrm{inf}$ & Ref &	$\dot{M}$ & Ref  \\
 &  &  & (K)  &  & (cgs) & & & & & & & & (km~s$^{-1}$) & &  (km~s$^{-1}$) & & ($M_{\odot}~$yr$^{-1}$) &  \\  
\hline
\endfirsthead
\multicolumn{4}{c}%
{\tablename\ \thetable\ -- \textit{Continued from previous page}} \\
\hline
Target &	SpType & Ref &$T_\mathrm{eff}$ & Ref &	$\log~g$ &	Ref & $\log~ L/L_{\odot}$ & Ref &	$R/R_{\odot}$ &	Ref & $M/M_{\odot}$ & Ref &	\vsini & Ref &	$v_\mathrm{inf}$ & Ref &	$\dot{M}$ & Ref  \\
 &  &  & (K)  &  & (cgs) & & & & & & & & (km~s$^{-1}$) & &  (km~s$^{-1}$) & & ($M_{\odot}~$yr$^{-1}$) &  \\  
\hline
\endhead
\hline \multicolumn{4}{r}{\textit{Continued on next page}} \\
\endfoot
\hline
\endlastfoot
\multicolumn{19}{c}{\bf SMC targets} \\
2dFS 163	&	O8 Ib(f)	&	1	&	-	&	-	&	-	&	-	&	-	&	-	&	-	&	-	&	-	&	-	&	-	&	-	&	-	&	-	&	-	&	-	\\
AzV 6	&	O9 III	&	2	&	-	&	-	&	-	&	-	&	-	&	-	&	-	&	-	&	-	&	-	&	-	&	-	&	-	&	-	&	-	&	-	\\
AzV 14	&	O5 V	&	3	&	44000	&	3	&	4.0	&	3	&	5.82	&	3	&	14.2	&	3	&	74.0	&	3	&	261	&	4	&	2000	&	3	&	1.0E-07	&	3	\\
AzV 15	&	O6.5 II(f)	&	5	&	39000	&	6	&	3.6	&	6	&	5.83	&	6	&	18.0	&	From L,Teff	&	47.2	&	6	&	120	&	6	&	2050	&	6	&	1.1E-06	&	6	\\
AzV 16	&	sgB0[e|	&	7	&	-	&	-	&	-	&	-	&	-	&	-	&	-	&	-	&	-	&	-	&	-	&	-	&	-	&	-	&	-	&	-	\\
AzV 18	&	B2 Ia	&	2	&	19000	&	8	&	2.3	&	8	&	5.44	&	8	&	49.0	&	8	&	28.0	&	8	&	49	&	8	&	325	&	9	&	2.3E-07	&	8	\\
AzV 22	&	B5 Ia	&	2	&	14500	&	8	&	1.9	&	8	&	5.04	&	8	&	53.0	&	8	&	19.0	&	8	&	46	&	8	&	280	&	8	&	2.3E-07	&	8	\\
\end{longtable}
\begin{flushleft}
References: 
{\bf 1} \citet{2004MNRAS.353..601E}, 
{\bf 2} \citet{1997A&A...317..871L},
{\bf 3} \citet{2004ApJ...608.1001M},
{\bf 4} \citet{2009ApJ...700..844P},
{\bf 5} \citet{2000PASP..112.1243W},
{\bf 6} \citet{2021A&A...647A.134B},
{\bf 7} \citet{1998A&A...340..117L},
{\bf 8} \citet{2004A&A...417..217T},
{\bf 9} \citet{2004ApJ...607..451E},
{\bf 10} \citet{2003MNRAS.338..360F},
{\bf 11} \citet{2015A&A...581A..21H},
{\bf 12} \citet{2013ApJ...763..101L},
{\bf 13} \citet{2016ApJ...817..113L},
{\bf 14} \citet{1998MNRAS.300..828P},
{\bf 15} \citet{2002ApJS..141..443W},
{\bf 16} \citet{2004ApJ...610.1021E},
{\bf 17} \citet{2006ApJ...638..409H},
{\bf 18} \citet{2006A&A...457..949M},
{\bf 19} \citet{2006A&A...456.1131M},
{\bf 20} \citet{2005A&A...434..677T},
{\bf 21} \citet{1995ApJ...438..188M},
{\bf 22} \citet{2013A&A...555A...1B},
{\bf 23} \citet{2006A&A...456..623E},
{\bf 24} \citet{2007A&A...471..625T},
{\bf 25} \citet{1987AJ.....93.1070G},
{\bf 26} \citet{2005ApJ...627..477M},
{\bf 27} \citet{2014ApJ...785..151H},
{\bf 28} \citet{2007A&A...466..277H},
{\bf 29} \citet{1995PASP..107..104W},
{\bf 30} \citet{2004ApJ...608.1028W},
{\bf 31} \citet{2009ApJ...692..618M},
{\bf 32} \citet{2019A&A...626A..50D},
{\bf 33} \citet{2008A&A...479..541H},
{\bf 34} \citet{2016A&A...591A..22S},
{\bf 35} \citet{2012A&A...537A..29R},
{\bf 36} \citet{1977ApJ...215...53W},
{\bf 37} \citet{1983ApJ...265..716W},
{\bf 38} \citet{1996A&A...305..171P},
{\bf 39} \citet{2018ApJ...868...57C},
{\bf 40} \citet{1982PASP...94...31C},
{\bf 41} \citet{2005A&A...436..687H},
{\bf 42} \citet{2003A&A...400...21R},
{\bf 43} \citet{2010AJ....139.1283W},
{\bf 44} \citet{1997AJ....114.1951S},
{\bf 45} \citet{2018A&A...616A.103S},
{\bf 46} \citet{2005A&A...434.1125D},
{\bf 47} \citet{2006A&A...451..603D},
{\bf 48} \citet{2012A&A...543A..95R},
{\bf 49} \citet{2019A&A...625A.104R},
{\bf 50} \citet{1991MNRAS.250..119B},
{\bf 51} \citet{1969AJ.....74..877S},
{\bf 52} \citet{1991PASP..103.1123F},
{\bf 53} \citet{2017AJ....154..102U},
{\bf 54} \citet{2004ApJ...617.1316P},
{\bf 55} \citet{2002ApJS..141...81M},
{\bf 56} \citet{1988ApJ...335..703F},
{\bf 57} \citet{1978A&AS...31..243R},
{\bf 58} \citet{1996MNRAS.281..163S},
{\bf 59} \citet{2014A&A...565A..27H},
{\bf 60} \citet{2007A&A...465.1003M},
{\bf 61} \citet{2012A&A...537A..79R},
{\bf 62} \citet{2002AJ....123.2754W},
{\bf 63} \citet{1992AJ....103.1205P},
{\bf 64} \citet{2017ApJ...837..122M},
{\bf 65} \citet{2000AJ....119.2214M},
{\bf 66} \citet{2002A&A...392..653C},
{\bf 67} \citet{2011MNRAS.416.1311C},
{\bf 68} \citet{2002ApJ...579..774C},
{\bf 69} \citet{1986AJ.....92...48C},
{\bf 70} \citet{2009PASP..121..634B},
{\bf 71} \citet{1987PASP...99..240G},
{\bf 72} \citet{2003ApJ...586..996M},
{\bf 73} \citet{2008MNRAS.389..806S},
{\bf 74} \citet{1997A&A...320..500C},
{\bf 75} Crowther (2021, priv.\ com.),
{\bf 76} \citet{2003MNRAS.338..141O},
{\bf 77} \citet{1998AJ....116.1263O},
{\bf 78} \citet{2018A&A...615A..40R},
{\bf 79} \citet{2018A&A...609A...7R},
{\bf 80} \citet{2015A&A...584A...5E},
{\bf 81} \citet{2002A&A...385..517N},
{\bf 82} \citet{2020A&A...634A.118M},
{\bf 83} \citet{2014A&A...564A..40W},
{\bf 84} \citet{2014A&A...570A..38B},
{\bf 85} \citet{2013A&A...560A..29R},
{\bf 86} \citet{2017A&A...600A..81R},
{\bf 87} \citet{2017A&A...601A..79S},
{\bf 88} \citet{2013A&A...560A..29R},
{\bf 89} \citet{2019A&A...627A.151S},
{\bf 90} \citet{2014A&A...564A..39S},
{\bf 91} \citet{2015A&A...574A..13E},
{\bf 92} \citet{2015A&A...575A..70M},
{\bf 93} \citet{2009AJ....138..510F},
{\bf 94} \citet{2018MNRAS.474.1412G},
{\bf 95} \citet{2004NewAR..48..727N},
{\bf 96} \citet{2007ApJ...659.1198E},
{\bf 97} \citet{2014A&A...572A..36T},
{\bf 98} \citet{2019MNRAS.484..422G},
{\bf 99} Garcia (2019, priv.\ com.). 
\end{flushleft}
\end{landscape}
\end{footnotesize}
\twocolumn


\bibliographystyle{aa}
\bibliography{bibliography,wg9}


\end{document}